\PassOptionsToPackage{noend}{algorithmic}

\documentclass{article}

\usepackage{hyperref}
\newcommand{\INDSTATE}[1][1]{\hspace{#1\algorithmicindent}}

\usepackage[accepted]{icml2020}

\icmltitlerunning{Causal Inference using Gaussian Processes with Structured Latent Confounders}

\usepackage{microtype}
\usepackage{graphicx}
\usepackage{caption}
\usepackage{booktabs} % for professional tables
\usepackage{amsmath}
\usepackage{mathtools}
\usepackage{subcaption}
\usepackage{csquotes}
\usepackage{amssymb}
\usepackage{amsthm}
\usepackage{multirow}
\usepackage{cuted}

\newtheorem{theorem}{Theorem}[section]

\newtheorem{prop}[theorem]{Proposition}

\hypersetup{draft}

\begin{document}

\twocolumn[
\icmltitle{Causal Inference using Gaussian Processes \\with Structured Latent Confounders}

\icmlsetsymbol{equal}{*}

\begin{icmlauthorlist}
\icmlauthor{Sam Witty}{UMass}
\icmlauthor{Kenta Takatsu}{UMass}
\icmlauthor{David Jensen}{UMass}
\icmlauthor{Vikash Mansinghka}{MIT}
\end{icmlauthorlist}

\icmlaffiliation{UMass}{College of Information and Computer Sciences, University of Massachusetts, Amherst, United States}

\icmlaffiliation{MIT}{Massachusetts Institute of Technology, Cambridge, United States}

\icmlcorrespondingauthor{Sam Witty}{switty@cs.umass.edu}

\icmlkeywords{Machine Learning, ICML}

\vskip 0.3in]

\printAffiliationsAndNotice{} 

\newcommand\inv[1]{#1\raisebox{1.15ex}{$\scriptscriptstyle-\!1$}}

\newcommand\U{\textbf{u}}

\newcommand\X{\mathrm{\textbf{x}}}

\newcommand\T{\textbf{t}}

\newcommand\Y{\textbf{y}}

\newcommand\eps{\boldsymbol{\epsilon}}

\newcommand\I{\textit{I}}
\newcommand\uniform{\text{Uniform}}
\newcommand\W{\textbf{w}}
\newcommand\K{K}
\newcommand\joint{P(Y, T, X, U, \Theta)}
\newcommand\ITE{\textit{ITE}}

\newcommand\SATE{\textit{SATE}}

\begin{abstract}
Latent confounders---unobserved variables that influence both treatment and outcome---can bias estimates of causal effects. In some cases, these confounders are shared across observations, e.g. all students taking a course are influenced by the course's difficulty in addition to any educational interventions they receive individually. This paper shows how to semiparametrically model latent confounders that have this structure and thereby improve estimates of causal effects. The key innovations are a hierarchical Bayesian model, \textit{Gaussian processes with structured latent confounders} (GP-SLC), and a Monte Carlo inference algorithm for this model based on elliptical slice sampling. GP-SLC provides principled Bayesian uncertainty estimates of individual treatment effect with minimal assumptions about the functional forms relating confounders, covariates, treatment, and outcome. 
%This paper also proves that, for linear functional forms, accounting for the structure in latent confounders is sufficient for asymptotically consistent estimates of causal effect.
Finally, this paper shows GP-SLC is competitive with or more accurate than widely used causal inference techniques %such as multi-level linear models and Bayesian additive regression trees. 
 on three benchmark datasets, including the Infant Health and Development Program and a dataset showing the effect of changing temperatures on state-wide energy consumption across New England.
\end{abstract}

\section{Introduction}
\label{sec:Introduction}

% Observational causal inference, the task of inferring the effects of interventions given non-randomized data, is central to a diversity of scientific and engineering disciplines. Unlike supervised machine learning, where the objective is to estimate the conditional distribution $P(Y|T)$ of some outcome variable $Y$ given a set of features $T$, the objective of causal inference is to estimate the distribution of $Y$ in the presence of some external manipulation on $T$, denoted $P(Y|do(T))$.

Multiple causal models can be \textit{observationally equivalent}, i.e., they induce the same likelihoods for observed data, while producing different estimates of the effects of an intervention. This observational equivalance between causal models is the basis for the colloquial expression \enquote{correlation does not imply causation.} Distinguishing between causal models, and estimating the effects of interventions, requires untestable assumptions about causal structure.

One such common assumption is \textit{unconfoundedness}~\cite{imbens2015causal}, i.e., that there exist no latent variables that influence both treatment and outcome. This assumption enables the unique identification of interventional distributions from the joint distribution over observed variables~\cite{Pearl2009Causality} and reduces causal inference to probabilistic estimation. Unfortunately, assuming unconfoundedness is often unreasonable in real observational settings~\cite{shadish2008can}. However, it may be more reasonable to assume uncounfoundedness for a subset of data instances that are known to share a common structure.

\begin{figure*}[t!]
    \centering
    \begin{subfigure}[b]{0.32\textwidth}
        \small{
           \begin{tabular}{c|c|c}
                Symbol & Description & Entity\\
                \hline
                $\U_o$ & Confounders & Object\\
                $\X_i$ & Covariates & Instance\\
                $\T_i$ & Treatment & Instance\\
                $\Y_i$ & Outcome & Instance
                \vspace{0.8cm}
            \end{tabular}
            }
        \caption{Variable descriptions.}
    \end{subfigure}
    ~
    \begin{subfigure}[b]{0.32\textwidth}
        \centering
        \includegraphics[width=0.8\textwidth]{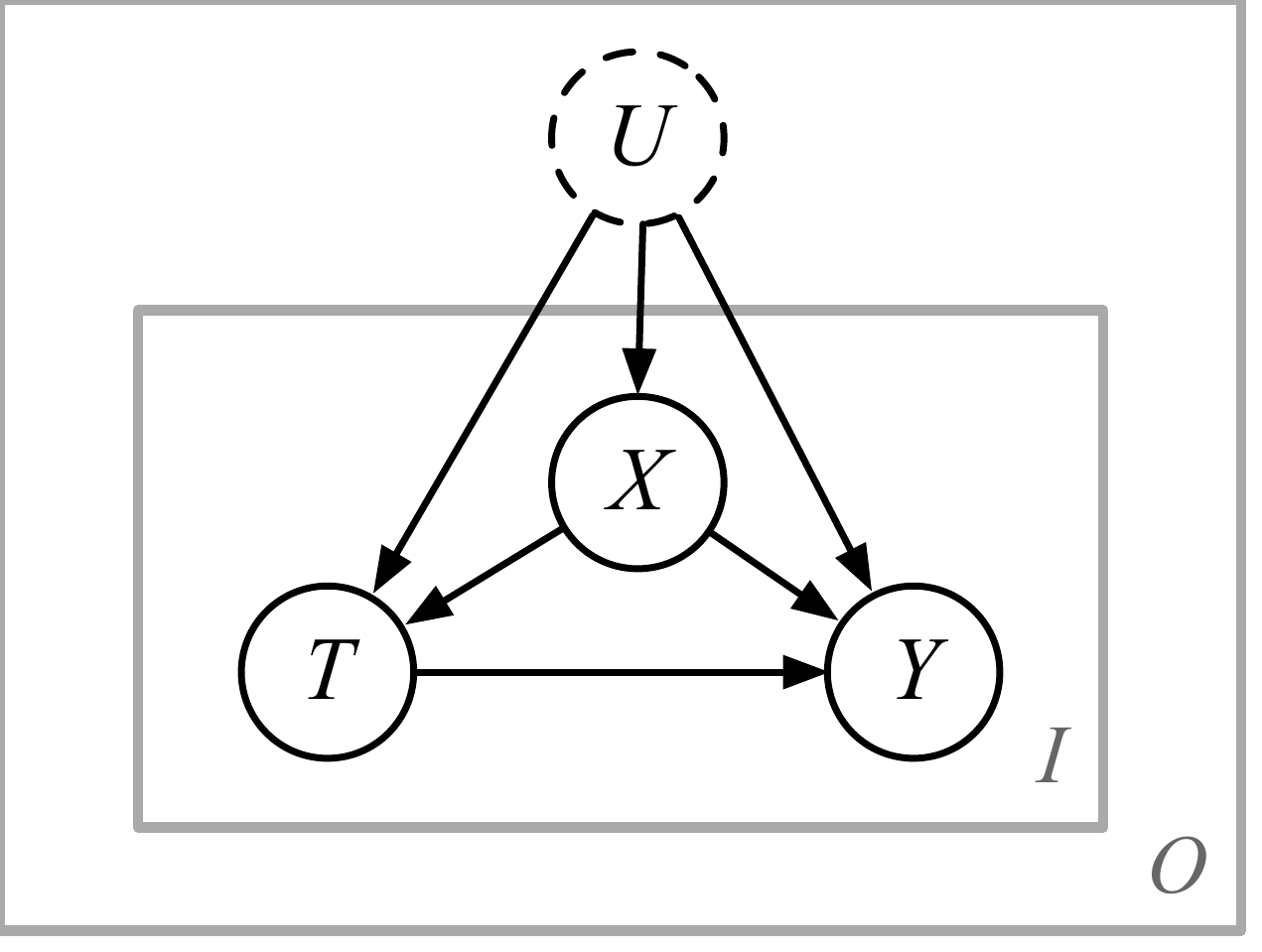}
        \caption{Causal graph for GP-SLC.}
        \label{fig:plate}
    \end{subfigure}
    ~
    \begin{subfigure}[b]{0.32\textwidth}
        \scriptsize{
        \begin{align*}
            f_u &\sim GP(0, k_u) &  f_x &\sim GP(0, k_x)\\ 
            f_t &\sim GP(0, k_t) & f_y &\sim GP(0, k_y)
        \end{align*}
        \vspace{-0.8cm}
        \begin{align*}
            \U_{o=1...N_O} &= f_u(\eps_{u_o})\\
            \X_{i=1...N_I} &= f_x(\U_{o=Pa(i)}, \eps_{x_i})\\
            \T_{i=1...N_I} &= f_t(\U_{o=Pa(i)}, \X_i, \eps_{t_i}) \\
            \Y_{i=1...N_I} &= f_y(\U_{o=Pa(i)}, \X_i, \T_i, \eps_{t_i})
        \end{align*}
        }
        \caption{Prior and causal functions for GP-SLC.}
    \end{subfigure}
    \\
    \begin{subfigure}[b]{0.32\textwidth}
        \centering
        \includegraphics[height=3cm]{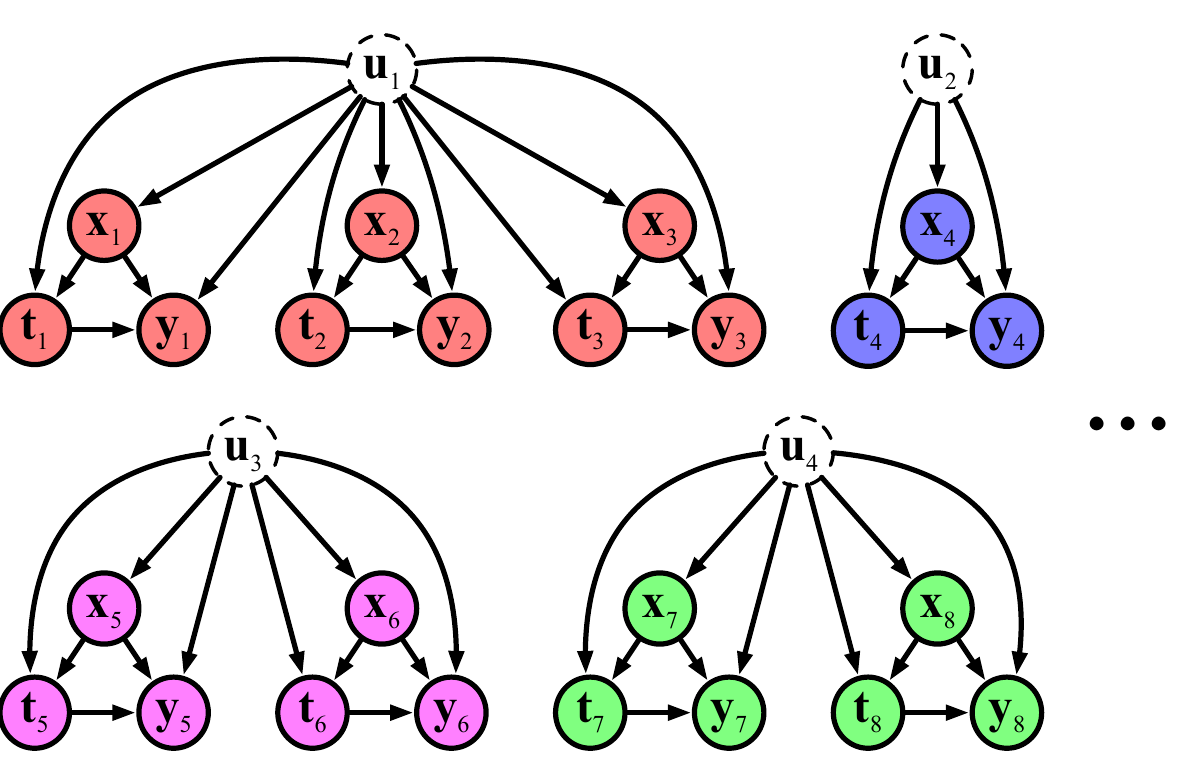}
        \caption{Example grounding of the structural causal model in (b) and (c). Latent confounders are shared within objects.}
    \end{subfigure}
    ~
    \begin{subfigure}[b]{0.32\textwidth}
        \centering
        \includegraphics[height=3cm]{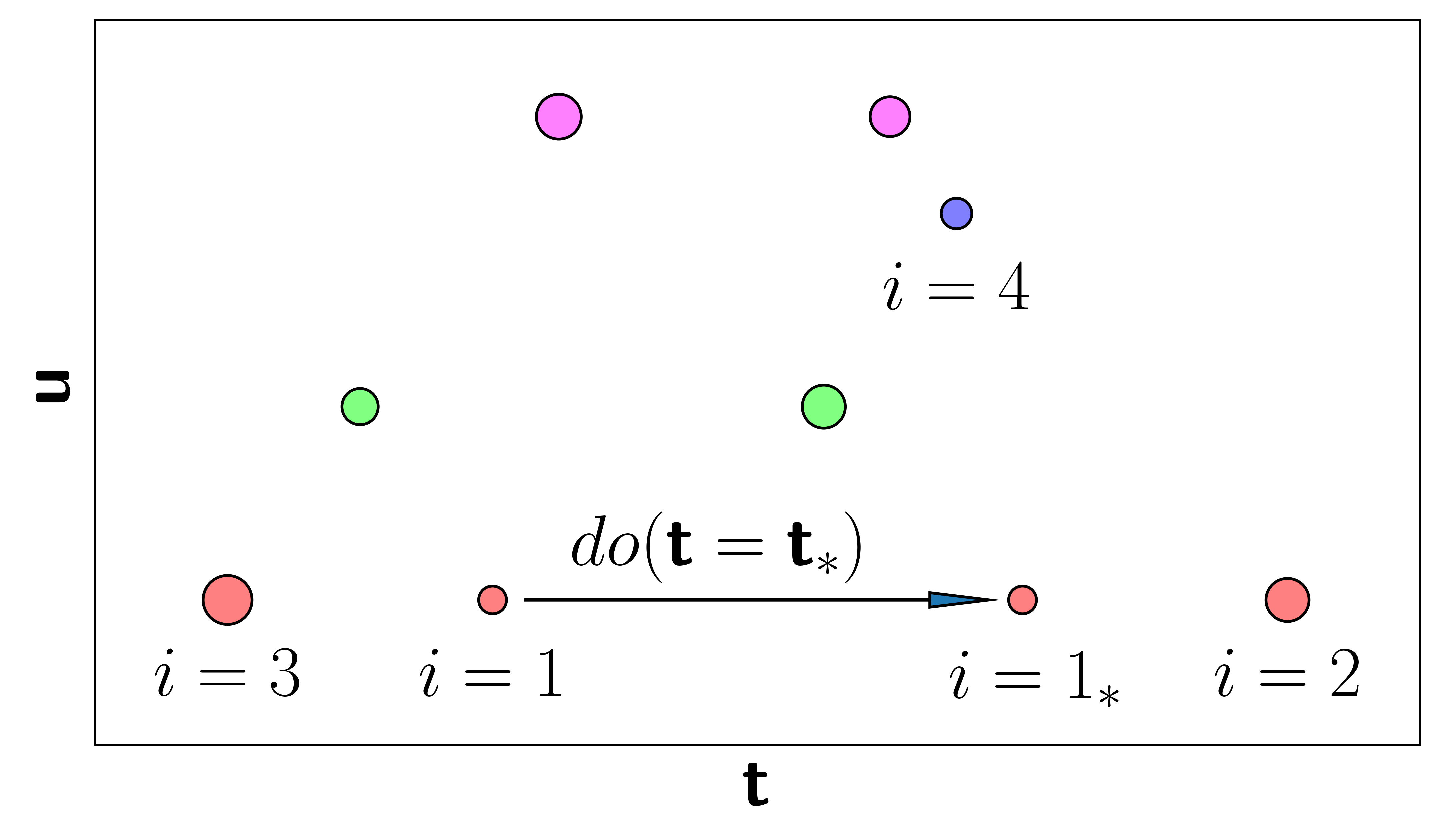}
        \caption{Treatment, covariates, and inferred object-level confounders for instances in (d). Color = $o$. Size = $\X$.}
    \end{subfigure}
    ~
    \begin{subfigure}[b]{0.32\textwidth}
        \centering
        \includegraphics[height=3cm]{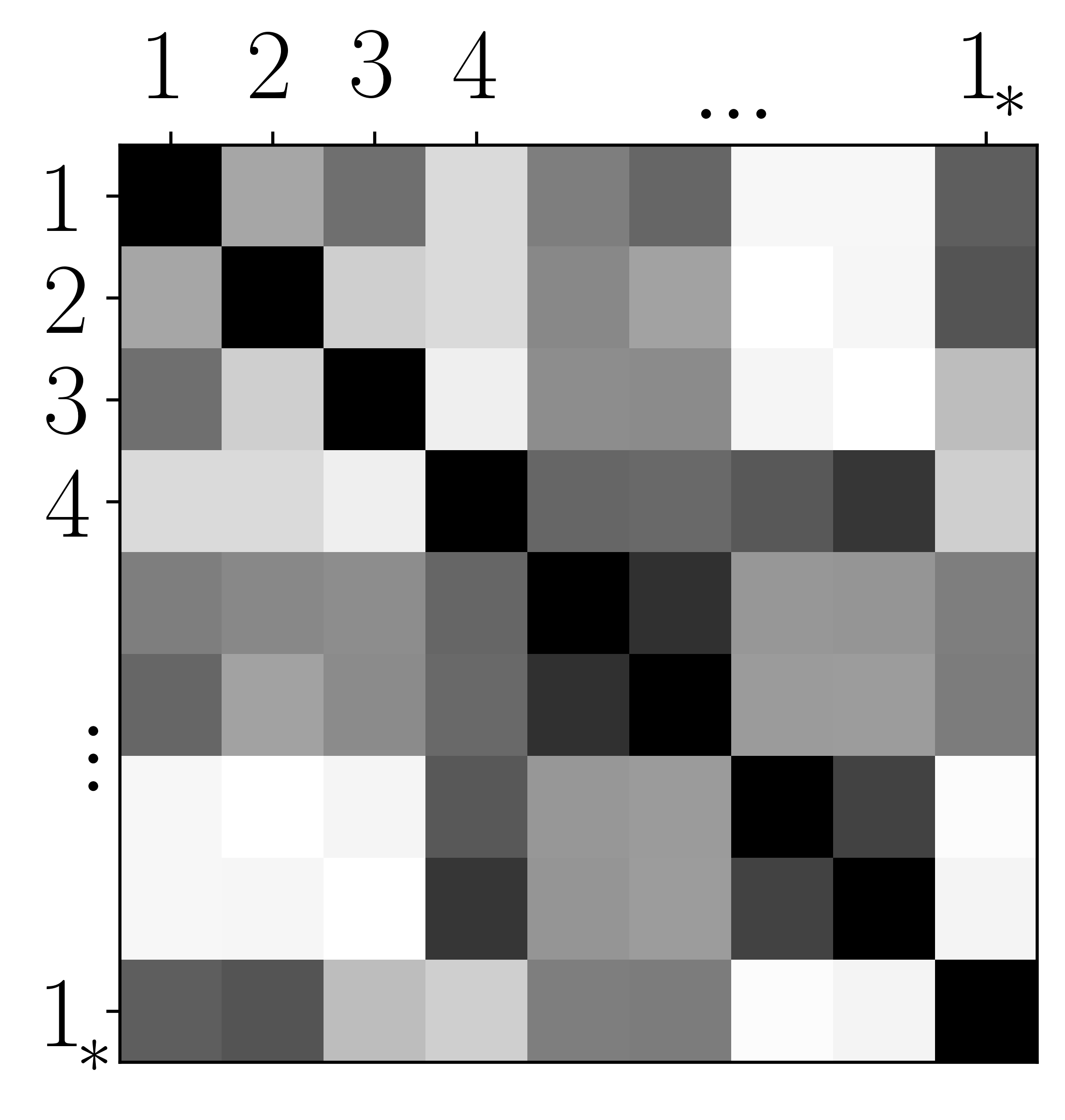}
        \caption{Kernel covariance matrix over observed ($\Y_i$) and counterfactual ($\Y_{1, t_*}$) outcomes for instances in (e). Dark $>$ light.}
    \end{subfigure}

    \caption{\textbf{Model summary.} GP-SLC (a-c) is a Gaussian process model for causal inference in settings where object-level latent confounders, $\U$, influence instance-level observed covariates, $\X$, treatment, $\T$, and outcome, $\Y$, random variables. For a given grounding (d), the outcome kernel function, $k_y$, applied to treatment, covariates, and inferred confounders (e) induces the covariance between observed and counterfactual outcomes (f). Instances belonging to the same object always have the same inferred latent $\U$. In this example, the counterfactual outcome $\Y_{1, t_*}$ has high covariance with factual outcomes $\Y_1$ and $\Y_2$. $\Y_{1, t_*}$ has low, but non-zero, covariance with $\Y_4$ because $\U_{Pa(1)} \not \approx \U_{Pa(4)}$, despite the fact that $\T_* \approx \T_4$ and $\X_1 \approx \X_4$.}
    \label{fig:intuition}
\end{figure*}
\raggedbottom

For example, suppose an educator proposes a new policy of holding back poor performing kindergarten students~\cite{hong2006evaluating, hong2008effects} with the intention of increasing their future academic performance. To estimate the effect of this policy change, they gather data on student retention and education outcomes from a national database. Here, the unconfoundedness assumption is not justified, as the schools' retention policies are likely to be influenced by local economic conditions, which may also influence student outcomes through other causal mechanisms, such as the availability of educational resources. However, the assumption may be justified when considering only students within a particular school, as this subset of students are similarly influenced by local economic conditions. In other words, statistical relationships within a school are less likely to be biased by latent confounders than are statistical relationships across the entire population. 

In this paper, we present Gaussian processes with structured latent confounders (GP-SLC), a novel Bayesian nonparametric approach to causal inference with hierarchical data. The key innovation behind GP-SLC is to place Gaussian process priors over functions in a hierarchical structural causal model, bringing the flexibility of Gaussian process models to a wide variety of practical causal inference techniques. GP-SLC naturally handles binary and continuous treatments and requires minimal assumptions about functional relationships between latent confounders, observed covariates, treatment, and outcomes. See Figure~\ref{fig:intuition} for an overview on how GP-SLC estimates counterfactual outcomes from data.

\section{Background}

\subsection{Object Conditioning}

Recent work has studied how the analytical procedure of partitioning data based on a known object hierarchy (e.g. students belonging to the same school) relates to the syntax and semantics of causal graphical models~\cite{jensen2019object}. This work concludes that conditioning on the identify of objects (referred to as \textit{object conditioning}) is distinct from existing notions of conditioning on the values of variables. Importantly, object conditioning constrains a set of latent variables to be identical across a set of instances, but does not constrain the particular value of those variables. Furthermore, the statistical implications of object conditioning differ from those of variable conditioning in that object conditioning does not induce collider bias when variables on the object are caused jointly by treatment and outcome. 

Partitioning hierarchical data in this way is the key analytical procedure for a variety of practical causal inference techniques, including within-subjects designs~\cite{loftus1994using}, difference-in-differences designs~\cite{shadish2008can}, longitudinal studies~\cite{liang1986longitudinal}, twin studies~\cite{boomsma2002classical}, and multi-level-modeling~\cite{gelman2006data}. As in the student retention example, these techniques take advantage of background knowledge about which instances (students) belong to which objects (schools) to mitigate the biasing effects of latent confounders. %\footnote{Such models also help to reduce variance in estimated effects given observed confounders.}
However, these methods typically rely on simple parametric assumptions, such as linear functional dependencies. These parametric assumptions are often unjustified in real domains, leading to poor estimates of causal effect.

We employ the idea of object conditioning directly in the GP-SLC model, constraining the joint distribution over individuals' latent confounders instead of treating object identity as a covariate in and of itself. By explicitly performing inference over object-level latent confounders, GP-SLC's estimates of counterfactual outcomes in one object are informed by observed outcomes in another. Sharing information between objects in this way is particularly valuable when each object contains few observed instances, as we show in Section~\ref{sec:Experiments}.

\subsection{Causal Inference with Latent Confounders}

Latent confounders---unobserved variables that cause both treatment and outcome---bias estimates of treatment effect. However, the effect can be adjusted for with additional background knowledge, such as that a latent confounder influences an observed proxy variable~\cite{kuroki2014measurement, miao2018proxy}. Similarly, recent work indicates that latent confounders can be adjusted for if they cause multiple candidate treatment variables~\cite{wang2019blessings}.

GP-SLC is similar to these approaches, in that it leverages additional background knowledge to adjust for latent confounders. However, unlike prior work using generative models for causal inference with latent confounders, it leverages known hierarchical structure to identify causal effects. 

\subsection{Gaussian Process Models}
\label{sec:GPR}

Gaussian process models are a flexible technique for probabilistic modeling. Specifically, a Gaussian process is a distribution over deterministic functions, $\Y=f(\X)$, $f \sim GP(m, k)$, which is fully specified by its mean function, $m(\X)$ and covariance function, $k(\X, \X')$, which we will refer to as the kernel function~\cite{rasmussen2003gaussian}. By definition, any finite collection of draws from a Gaussian process prior are jointly Gaussian distributed, $Y \sim \mathcal{N}(\mu, \Sigma)$, where $\mu_i = m(\X_i)$ and $\Sigma_{i,{i'}} = k(\X_i, \X_{i'})$. We denote such covariance matrices as $\K(X, X)$, where $X$ and $Y$ are matrices of all $\X_i$ and $\Y_i$ respectively. It is common to set the prior mean function to $m(\X)=0$, which we do in GP-SLC. 

This identity is useful for two reasons: (i) it provides an explicit likelihood, which can be used to perform inference over latent variables~\cite{lawrence2004gaussian, titsias2010bayesian}; and (ii) it enables closed-form out-of-sample probabilistic prediction~\cite{rasmussen2003gaussian}. We take advantage of both of these characteristics in GP-SLC, performing approximate inference over latent confounders in Section~\ref{subsec:approx} and predicting counterfactual outcomes in Section~\ref{subsec:exact}.

\subsection{Structural Causal Models}
\label{sec:CGMs}

GP-SLC can be thought of as a Bayesian nonparametric prior over functions in a structural causal model (SCM). SCMs provide a syntax and semantics for reasoning about interventional and counterfactual distributions in a system of random variables~\cite{Pearl2009Causality}. Counterfactuals~\cite{pearl2011algorithmization}---answers to what-if questions---are expressed in probability notation as $P(\Y_{\T_*}|\Y, \T)$, where $\Y_{\T_*}$ is the counterfactual outcome under intervention $do(\T=\T_*)$, $\Y$ is the observed outcome, and $\T$ is the observed treatment. In our education example, we may be interested in the counterfactual, \enquote{given that the student was not retained in kindergarten and they performed poorly in high school, how would they have performed if they had been retained?} Answering these kinds of counterfactual queries involves: (i) computing the conditional distribution over latent variables and exogenous noise given observed evidence; (ii) applying the intervention to the structural causal model; and (iii) recomputing the distribution over the outcome variable(s) using the modified structural causal model. We apply this procedure to estimate counterfactual outcomes using GP-SLC in Section~\ref{sec:GP-SLC}.

\section{Gaussian Processes with Structured Latent Confounders}
\label{sec:GP-SLC}

Consider the common scenario where there are $N_O$ object-level latent confounders $(\mathit{\U_o \in \mathbb{R}^{N_U}})$ that influence $N_I$ instances of observed treatment $(\T_i \in \mathbb{R})$, covariates $(\X_i \in \mathbb{R}^{N_X})$, and outcomes ($\Y_i \in \mathbb{R}$). We can describe this scenario as a structural causal model, where the particular functions relating $\U$, $\X$, $\T$, and $\Y$ are given by the following:

\vspace{-0.5cm}
\begin{gather}
    \begin{aligned}
        \U_{o=1...N_O} &= f_{u}(\eps_{u_o})\\
        \X_{i=1...N_{I}} &= f_{x}(\U_{o=Pa(i)}, \eps_{x_{i}})\\
        \T_{i=1...N_{I}} &= f_t(\U_{o=Pa(i)}, \X_i, \eps_{t_i}) \\
        \Y_{i=1...N_{I}} &= f_y(\U_{o=Pa(i)}, \X_i, \T_i, \eps_{y_i}).
    \end{aligned}
\label{eq:functional_causal}
\end{gather}

If all instances belong to the same object ($N_O = 1$) the structural causal model in Equation~\ref{eq:functional_causal} reduces to the standard propositional case and the latent $\U$ will not bias estimated counterfactual outcomes. However, if we wish to estimate counterfactual outcomes using instances from multiple objects ($N_I > N_O > 1$), $\U$'s influence on $\T$ and $\Y$ would appear to render counterfactual queries unidentifiable~\cite{Pearl2009Causality}. To address this problem of identifiability, GP-SLC places a mean-zero Gaussian process prior over each function in the structural causal model in Equation~\ref{eq:functional_causal}, with kernel functions $k_x$, $k_t$, and $k_y$ respectively as follows:

\vspace{-0.5cm}
\begin{gather*}
    \begin{aligned}
    f_{x} &\sim GP(0, k_{x}) & f_t &\sim GP(0, k_t) & f_y &\sim GP(0, k_y).
    \end{aligned}
\end{gather*}

The particular choice of each kernel function plays an important role in the prior over functions, and by extension the conditional distribution over counterfactual outcomes. We use a radial basis function (RBF) kernel with automatic relevance determination (ARD)~\cite{neal2012bayesian} and additive Gaussian exogenous noise for each Gaussian process prior. Each kernel is parameterized by a set of kernel lengthscales, $\lambda$, scaling factors, $\sigma^2$, and exogenous noise variances $\sigma^2_\epsilon$. We assume $f_u$ is the identity function. We refer to the noise-free component of each kernel function as $k'$, e.g. $k_{t}= k'_{t}([\U_{o=Pa(i)}, \X_i], [\U_{o'=Pa(i')}, \X_{i'}]) + \sigma^2_{\epsilon_y} \delta_{i, i'}$, where $\sigma^2_{\epsilon_y}$ is the exogenous noise variance, $\delta_{i, i'}$ is the Dirac-delta function at $i' = i$, and $k'_{t}$ is the ARD kernel. See the supplementary materials for detailed descriptions of these kernels.

In addition to placing Gaussian process priors on the functions in the structural causal model in Equation~\ref{eq:functional_causal}, we also place inverse-gamma priors, $P(\theta) = \inv{\gamma}(\theta; \alpha_{\theta}, \beta_{\theta})$ on each $\theta \in \Theta$, where $\Theta$ is the set of all kernel lengthscales, scaling factors, and exogenous noise variances. In Section~\ref{subsec:approx} we show how to perform approximate posterior inference on $\Theta$.

\subsection{Conditional Density}
As $f_y, f_t, \text{ and } f_x$ are all drawn from Gaussian process priors, $P(Y|T, X, U, \Theta), P(T| X, U, \Theta), \text{ and } P(X|U, \Theta)$ are all multivariate Gaussian distributions with mean zero and covariance given by their respective kernel covariance matrices. For example, $P(T|X, U, \Theta) =$ $ \mathcal{N}(T; 0, \K_t([U, X], [U, X]))$, where $\K_t([U, X], [U, X])_{i,i'}$ $= k_t([\U_{o=Pa(i)}, \X_i], [\U_{o'=Pa(i')}, \X_{i'}])$. As $\U_o$ is given by the identity function of exogenous Gaussian noise,  $P(\U_o|\Theta)$ $= \mathcal{N}(\U_o; 0, \sigma^2_{\epsilon_{u}}\I_{N_U})$. Therefore, the joint density is given by the following, which we use in Algorithms~\ref{alg:inference_hyp} and \ref{alg:inference_conf}:

\begin{algorithm}[t]
   \caption{Individual Treatment Effect Estimation}
   \label{alg:inference}
\begin{algorithmic}[1]
   \STATE {\bfseries Input:} 
   \STATE \INDSTATE Intervention assignment: $\T_*$
   \STATE \INDSTATE Data: $Y, T, X$
   \STATE \INDSTATE Prior hyperparameters: $\alpha_{\theta \in \Theta}, \beta_{\theta \in \Theta}$
   \STATE \INDSTATE Inference parameters: $\textrm{N}_\textrm{Outer}, \textrm{N}_\textrm{MH}, \textrm{N}_\textrm{ES}, \textrm{drift}_{\theta \in \Theta}$
   \STATE {\bfseries Procedure:}
   \begin{ALC@g}
   \STATE $\theta \sim \inv{\gamma}(\alpha_{\theta}, \beta_{\theta}), \forall \theta \in \Theta$ \, \; \; \; \; \; \; \; \; $\triangleright$ Prior sample
   \STATE $\U_{o} \sim \mathcal{N}(0, \sigma^2_{U} \I_{N_U}), \forall o=1...N_O$ \; \; $\triangleright$ Prior sample
   \STATE $\textit{ITESamples} \gets \{ \null \}$
  \FOR{$l=1$ {\bfseries to} $n_{Outer}$}
  \STATE $\Theta \gets \text{HyperparameterUpdate}(...)$ \, $\triangleright$ Algorithm \ref{alg:inference_hyp}
  \STATE $U \gets \text{ConfounderUpdate}(...)$ \; \; \; \; $\triangleright$ Algorithm \ref{alg:inference_conf}
  \STATE $\W_i \gets [\T_i, \X_i, \U_{o=pa(i)}], \forall i \in 1...N_I$
  \STATE $\W_{i, *} \gets [\T_*, \X_i, \U_{o=pa(i)}], \forall i \in 1...N_I$
  \STATE $\mu_{\ITE} \gets (\K'(W, W_*) \text{-} \K'(W, W)) \inv{\K(W, W)}Y$ 
  \STATE $\ITE \sim \mathcal{N}(\mu_{\ITE}, \Sigma_{\ITE})$ $\triangleright$ See Supplement for $\Sigma_{\ITE}$
  \STATE $\textit{ITESamples} \gets \ITE\cup \textit{ITESamples}$
  \ENDFOR
  \RETURN $\textit{ITESamples}$
  \end{ALC@g}
  \end{algorithmic}
\end{algorithm}
\raggedbottom

\vspace{-0.6cm}
\begin{gather*}
\begin{aligned}
    \joint{} =& P(Y|T, X, U, \Theta)P(X|U, \Theta)\\ 
    & P(T|X, U, \Theta) \prod_{o=1...N_O} P(\U_o|\Theta)P(\Theta).
\end{aligned}
\end{gather*}

By placing Gaussian process priors over each function in the hierarchical structural model, we encode our assumptions about which configurations of observed and latent variables are reasonable a-priori. Using a radial basis function kernel, we assume that if two objects have similar object-level latent confounders, they are likely to induce similar distributions over observed covariates, treatment, and outcome. Placing higher density on smooth structural causal functions in this way enables inference over object-level confounders.

\section{Estimating Treatment Effects}
\label{sec:Estimation}

In this section we describe how to estimate the \textit{individual treatment effect}, $\ITE_{i, t_*} = \Y_{i, t_*} - \Y_i$, the difference between observed and counterfactual outcomes for the $i$th instance. Standard aggregate measures of causal effect, such as the \textit{sample average treatment effect}, $\SATE_{t_*} = \frac{1}{N_I}\sum_{i} \ITE_{i, t_*}$, can be derived from the individual treatment effect. We use $\ITE_{t_*}$ to denote the vector of individual treatment effects for the intervention $do(\T_i = \T_*)$ applied uniformly to each instance $i$, although the estimation procedure can be easily applied to any arbitrary set of intervention assignments.

First, note that when exogenous noise is additive in $f_y$, i.e $f_y(\U_{o=Pa(i)}, \X_i, \T_i, \eps_{t_i}) = $ $f'_y(\U_{o=Pa(i)}, \X_i, \T_i) + \eps_{t_i}$, as in the GP-SLC model, individual treatment effect is given by the difference between noise-free functions $\ITE_{i, t_*} = f'_y(\U_{o=Pa(i)}, \X_i, \T_*) - f'_y(\U_{o=Pa(i)}, \X_i, \T_i)$. We denote the outcome of these noise-free functions as $\Y'_{i, t_*}$ and $\Y'_i$, and the vector of outcomes as $Y'_{t_*}$ and $Y'$ respectively.\footnote{Noise-free prediction is often denoted as $f$ in Gaussian process regression models. We avoid this notation to avoid confusion with functions in the structural causal model.} As $U \cup X$ blocks all backdoor paths from $T$ to $Y$, we have that the distribution over individual treatment effects is given by the following expression~\cite{Pearl2009Causality}:
\vspace{-0.1cm}
\begin{gather*}
\begin{aligned}
    &P(\ITE_{t_*}|Y, T, X) = P(Y'_{t_*} - Y'|Y, T, X)\\ 
    &=\int P(Y'_{*}-Y'|T_*, Y, T, X, U, \Theta) P(U, \Theta|Y, T, X) dU d\Theta.
\end{aligned}
\end{gather*}

This equation directly informs our hybrid procedure for estimating counterfactual outcomes shown in Algorithm~\ref{alg:inference}, (i) generate approximate samples from the posterior $\hat{U}, \hat{\Theta} \sim P(U, \Theta|Y, T, X)$ and (ii) for each posterior sample compute the conditional distribution $(Y'_*-Y'|T_*, Y, T, X, \hat{U}, \hat{\Theta})$ in closed-form, taking advantage of Gaussian closure under conditioning and subtraction. As the posterior distribution $P(U, \Theta|Y, T, X)$ is intractable for non-trival kernels, we turn to Monte Carlo approximate inference techniques.

\subsection{Approximate Inference: $U$ and $\Theta$}
\label{subsec:approx}

Because we assume that our structural functions were drawn from Gaussian Processes, which provide a closed-form expression for the conditional density of the data, we are able to use standard likelihood-based approximate inference techniques. In our experiments, we approximate this posterior distribution using elliptical slice sampling~\cite{murray2010elliptical} for the latent confounder, $U$, and random walk Metropolis Hastings~\cite{hastings1970monte} on all kernel hyperparameters and exogenous noise variances, $\Theta$. Psuedo-code implementations are presented in Algorithms~\ref{alg:inference_hyp} and \ref{alg:inference_conf}. 
% \hspace{-3cm}

\subsection{Exact Inference: $Y'_* - Y'$}
\label{subsec:exact}
To estimate $P(Y'_*-Y'|T_*, Y, T, X, U, \Theta)$, we extend the Gaussian process model over in-sample and out-of-sample outcomes~\cite{rasmussen2003gaussian}. Using the shorthand $\W_i = [\T_i, \X_i, \U_{o=pa(i)}]$ and $\W_{i, *} = [\T_{*}, \X_i, \U_{o=pa(i)}]$, the joint distribution over observed outcomes, $Y$, noise-free outcomes for each observed instance, $Y'$, and noise-free counterfactual outcomes, $Y'_*$ conditioned on observed treatments, $T$, covariates, $X$, inferred confounders, $U$, and kernel hyperparameters, $\Theta$, is Gaussian distributed as follows, where $\K(W, W)=\K'(W, W) + \sigma^2_Y \I_{N_I}$ and $\K'(W, W)$ is the kernel matrix of $k'_y$ given $\Theta$:

\vspace{-0.5cm}
\begin{align*}
% \label{eq:exact_joint}
    \Bigg(\begin{bmatrix*}[l]Y \\Y' \\ Y'_*\end{bmatrix*}&|T_*, T, X, U, \Theta \Bigg) \\
    \sim \mathcal{N}& \Bigg(0, \begin{bmatrix*}[l] \K(W, W) & \K'(W, W) & \K'(W, W_*) \\ \K'(W, W) & \K'(W, W) & \K'(W, W_*) \\ \K'(W_*, W) & \K'(W_*, W) & \K'(W_*, W_*) \end{bmatrix*} \Bigg).
\end{align*}

As Gaussian distributions are closed under conditioning and subtraction, we have that $(Y'_*-Y'|T_*, Y, T, X, U, \Theta)$ is also jointly Gaussian distributed as follows, where $\mu_{\ITE} = (\K'(W, W_*) - \K'(W, W)) \inv{\K(W, W)}Y$:

\vspace{-0.4cm}
\begin{gather}
    \begin{aligned}
        (Y'_*-Y'|&T_*, Y, T, X, U, \Theta) \sim \mathcal{N}(\mu_{\ITE}, \Sigma_{\ITE})
    \end{aligned}
    % \label{eq:exact_conditional}
\end{gather}

See the supplementary materials for details and for a closed-form expression for $\Sigma_{\ITE}$.

\begin{algorithm}[t]
  \caption{Hyperparameter Update - Random Walk MH}
  \label{alg:inference_hyp}
\begin{algorithmic}[1]
  \STATE {\bfseries Input:} 
  \STATE \INDSTATE Data: $Y, T, X$
  \STATE \INDSTATE Posterior sample: $U, \Theta$
  \STATE \INDSTATE Prior hyperparameters: $\alpha_{\theta \in \Theta}, \beta_{\theta \in \Theta}$
  \STATE \INDSTATE Inference parameters: $\textrm{N}_\textrm{MH}, \textrm{drift}_{\theta \in \Theta}$
  \STATE {\bfseries Procedure:}
  \begin{ALC@g}
        \FOR{$j=1$ {\bfseries to} $\textrm{N}_\textrm{MH}$}
            \FOR{$\theta \in \Theta$}
                % Justify why we do this. Sample from an inverse gamma with mean = prev sample's \theta and a fixed variance.
                \STATE $\alpha_{\theta'} \gets \theta^2/\textrm{drift}_{\theta}$
                \STATE $\beta_{\theta'} \gets \theta(\alpha_{\theta'} - 1)$
                \STATE $\theta' \sim \inv{\gamma}(\alpha_{\theta'}, \beta_{\theta'})$
                \STATE $\alpha_{\theta''} \gets \theta'^2/\textrm{drift}_{\theta'}$
                \STATE $\beta_{\theta''} \gets \theta'(\alpha_{\theta''} - 1)$
                \STATE $\Theta' \gets \Theta \setminus \theta \cup \theta'$
                \STATE $A \gets \dfrac{P(Y, T, X, U, \Theta')}{P(Y, T, X, U, \Theta)} \dfrac{\inv{\gamma}(\theta';\alpha_{\theta'}, \beta_{\theta'})}{\inv{\gamma}(\theta;\alpha_{\theta''}, \beta_{\theta''})}$
                \STATE $\eta \sim \uniform(0, 1)$
                \IF{$\eta > \min(A, 1)$}
                    \STATE $\Theta = \Theta'$
                \ENDIF
            \ENDFOR
        \ENDFOR
        \RETURN $\Theta$
    \end{ALC@g}
\end{algorithmic}
\end{algorithm}
\raggedbottom

\begin{algorithm}[t]
  \caption{Confounder Update - Elliptical Slice Sampling}
  \label{alg:inference_conf}
\begin{algorithmic}[1]
  \STATE {\bfseries Input:} 
  \STATE \INDSTATE Data: $Y, T, X$
  \STATE \INDSTATE Posterior sample: $U, \Theta$
  %\STATE \INDSTATE Prior hyperparameters: $\alpha_{\theta \in \Theta}, \beta_{\theta \in \Theta}$
  \STATE \INDSTATE Inference parameter: $\textrm{N}_{\textrm{ES}}$
  \STATE {\bfseries Procedure:}
  \begin{ALC@g}
   % Indent what you need
   \FOR{$j=1$ {\bfseries to} $\textrm{N}_{\textrm{ES}}$}
            \FOR{$k=1$ {\bfseries to} $\textrm{N}_{\textrm{U}}$}
                \STATE $\textrm{done} \gets \textrm{False}$
                \STATE $\nu \sim \mathcal{N}(0, \sigma^2_{\epsilon_U} \I_{})$
                \STATE $y \sim \uniform(0, \joint)$
                \STATE $\phi \sim \uniform(0, 2\pi)$
                \STATE $[\phi_{\textrm{min}}, \phi_{\textrm{max}}] \gets [\phi-2\pi, \phi]$
                \WHILE{not $\textrm{done}$}
                    \STATE $U'_{*, k} \gets U_{*, k} \cos{\phi} + \nu \sin{\phi}$
                    \IF{$\joint > y$}
                        \STATE $\U_k \gets \U'_i$
                        \STATE $\textrm{done} \gets \textrm{True}$
                    \ELSE
                        \STATE \algorithmicif $\; \phi < 0$ \algorithmicthen $\; \phi_{\textrm{min}} \gets \phi$ \algorithmicelse $\; \phi_{\textrm{max}} \gets \phi$
                        \STATE $\phi \sim \uniform(\phi_{\textrm{min}}, \phi_{\textrm{max}})$
                    \ENDIF
                \ENDWHILE
            \ENDFOR
        \ENDFOR
        \RETURN $U'$
    \end{ALC@g}
\end{algorithmic}
\end{algorithm}
\raggedbottom

\section{Asymptotic Posterior Consistency}
\label{sec:Theory}

In the special case where each RBF kernel in the GP-SLC model is replaced with a linear kernel, $k(A, A') = A \cdot A'$, shared confounding among instances enables asymptotically consistent estimates of individual treatment effect. This is contrasted with the propositional setting (i.e. $N_O = N_I$) which does not lead to asymptotically consistent counterfactual estimation. Informally, a continuous random variable $\psi$ is asymptotically consistent if its posterior $P(\psi|data)$ approaches a Dirac-delta distribution at some point $\psi'$, regardless of the prior $P(\psi)$. We present proofs of Proposition~$\ref{prop:diag}$ and Theorems~\ref{thm:bigblocks} and \ref{thm:manyblocks} in the supplementary materials.

\begin{prop}
When $N_O = N_I$, $\ITE_{t_*}$ is not asymptotically consistent $\forall t_* \in \mathbb{R}$. 
\label{prop:diag}
\end{prop}

\begin{theorem}
\label{thm:bigblocks}
Assume there exists an object $o$ that is the parent of $n$ instances, $I' = \{i'_1, ..., i'_n\}$. Then $\ITE_{t_*}$ is asymptotically consistent as $n$ approaches $\infty, \forall t_* \in \mathbb{R}$.
\end{theorem}

\begin{theorem}
\label{thm:manyblocks}
Assume there exist $n$ objects $\mathbb{O} = \{o_1, ..., o_n\}$, each of which are the unique parents of $k \geq 2$ instances $I'_o = \{i'_{o,1}, ..., i'_{o, k_o}\}$. Then $\ITE_{t_*}$ is asymptotically consistent as $n$ approaches $\infty, \forall t_* \in \mathbb{R}$.
\end{theorem}

\begin{figure*}[t]
    \begin{subfigure}[b]{0.04\textwidth}
    \centering
    \includegraphics[height=2.6cm]{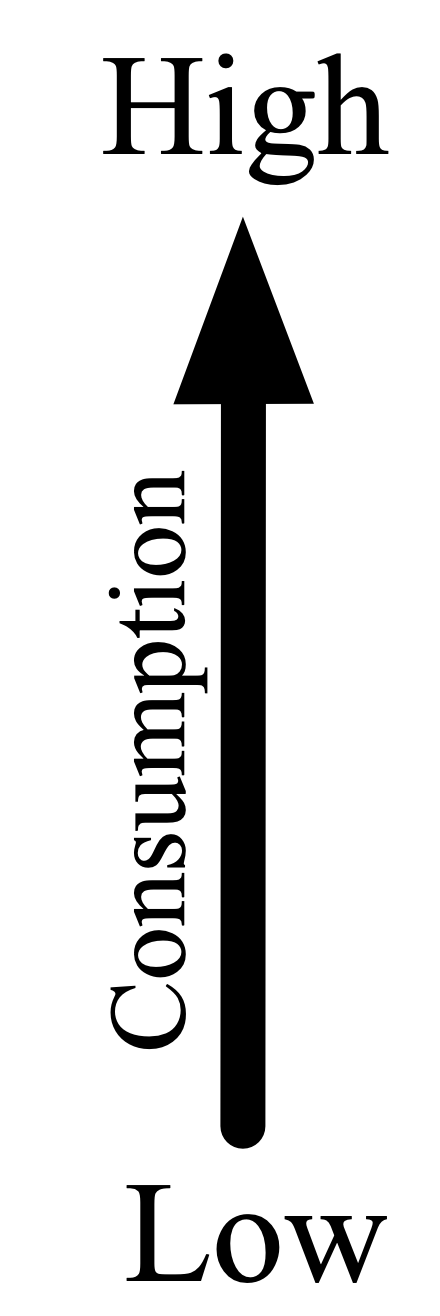}
    \vspace{-0.1cm}
    \end{subfigure}
    \begin{subfigure}[t]{0.2\textwidth}
        \centering
        \includegraphics[height=3cm]{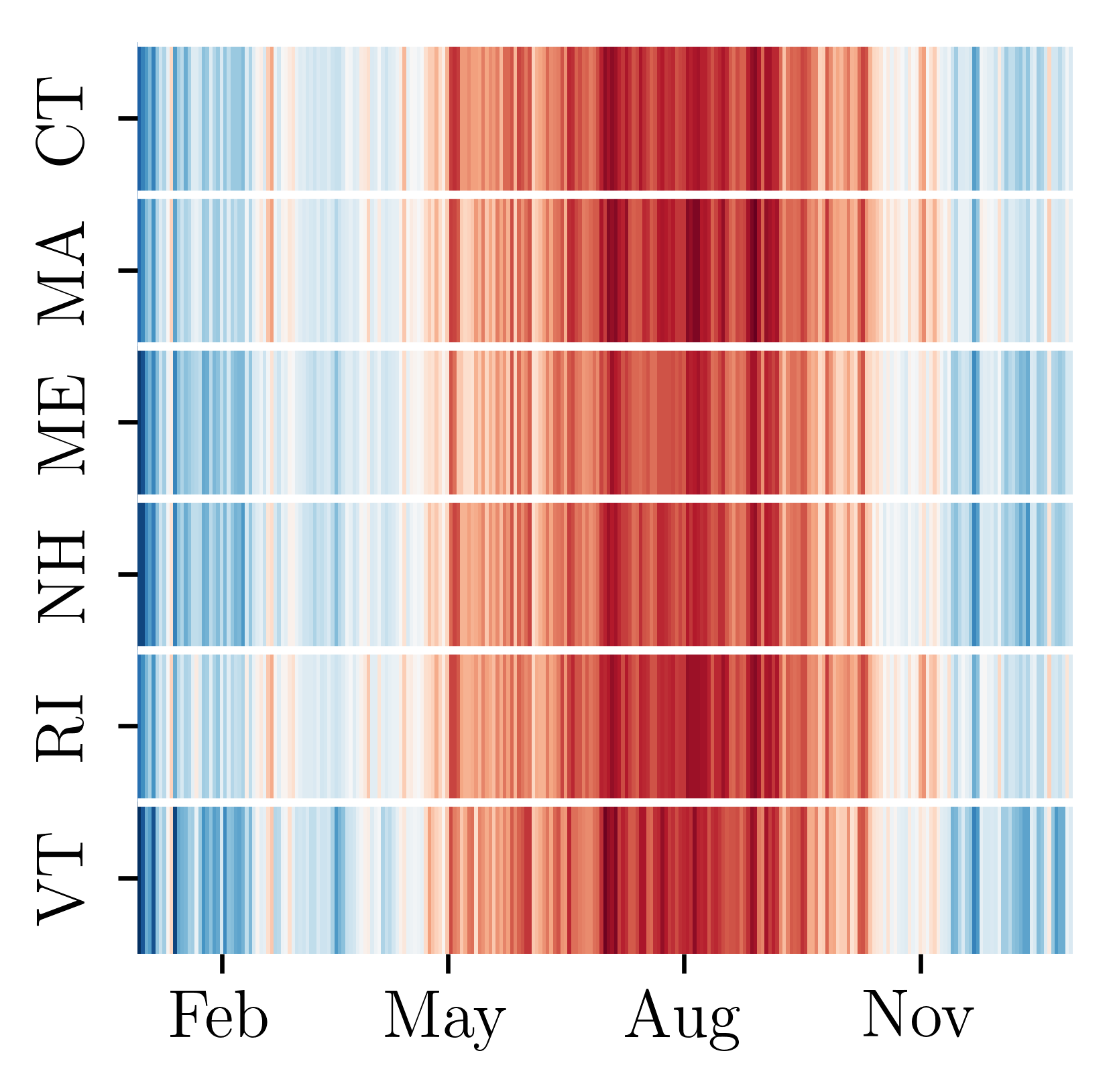}
        \caption{Original data.}
    \end{subfigure}%
    ~ 
    \hspace{-0.4cm}
    \begin{subfigure}[t]{0.2\textwidth}
        \centering
        \includegraphics[height=3cm]{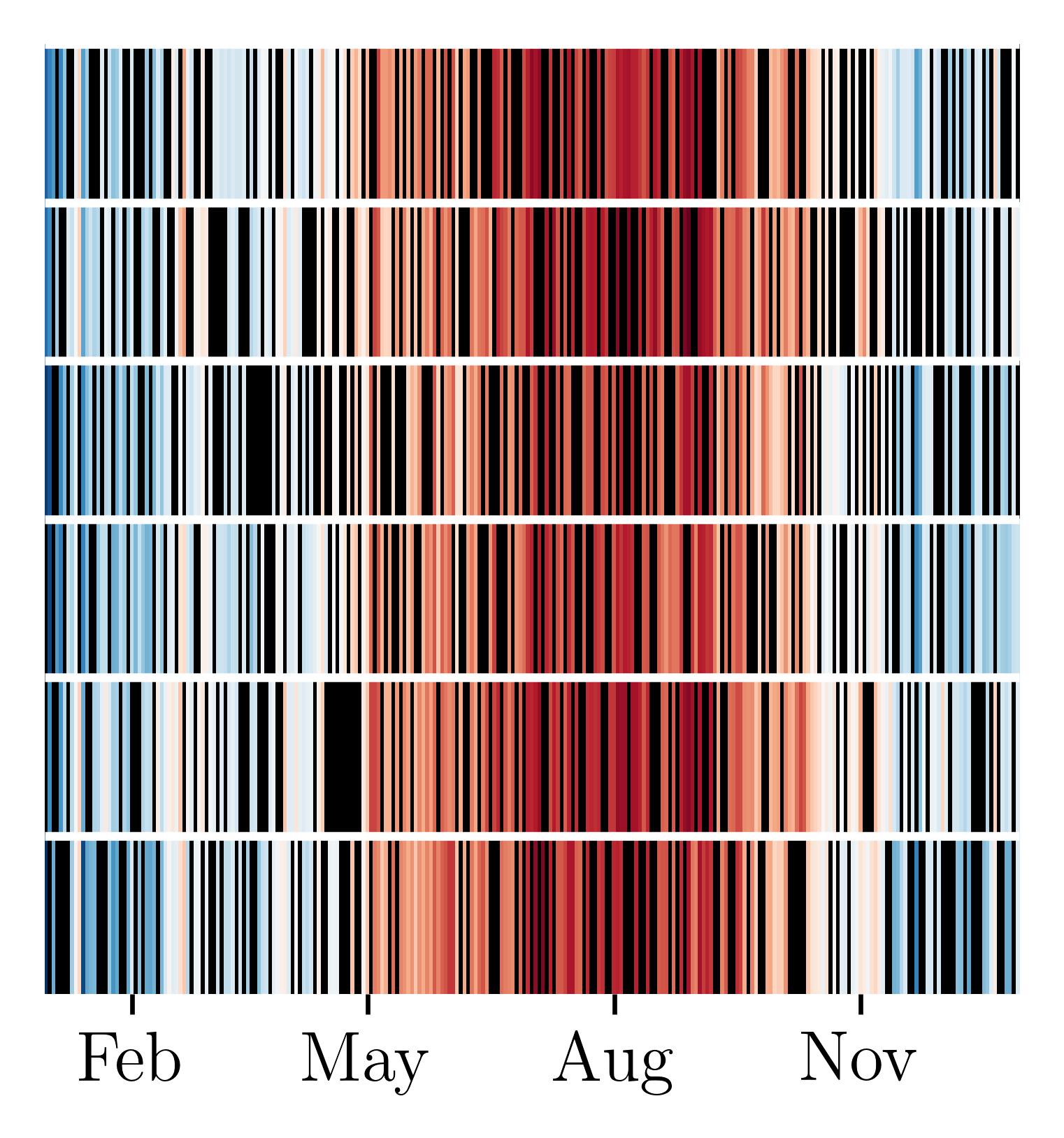}
        \caption{Unbiased sampling.}
    \end{subfigure}
    ~
    \hspace{-0.6cm}
    \begin{subfigure}[t]{0.2\textwidth}
        \centering
        \includegraphics[height=3cm]{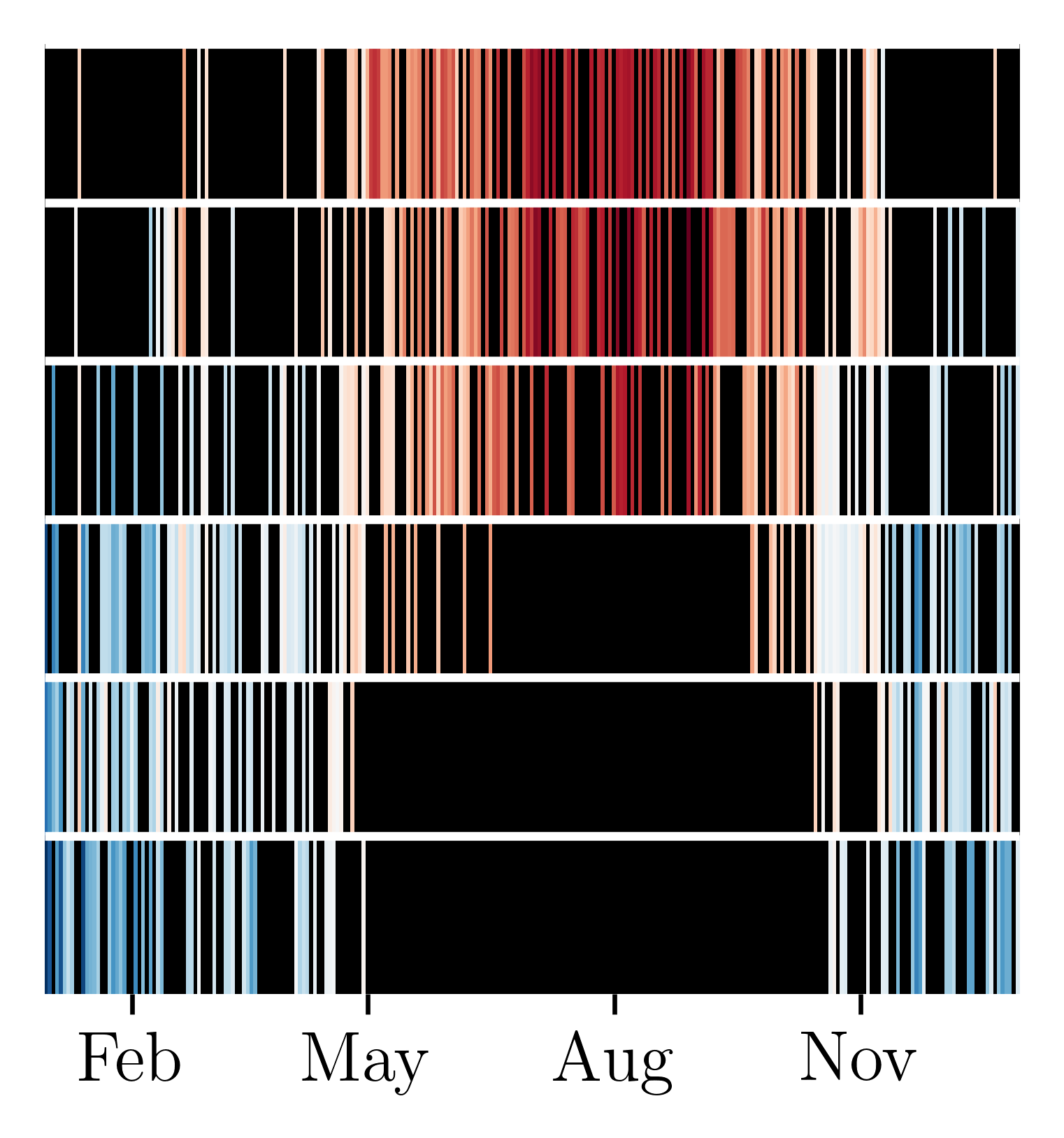}
        \caption{Biased sampling.}
    \end{subfigure}
    ~
    \hspace{-0.4cm}
    \begin{subfigure}[t]{0.05\textwidth}
        \centering
        \includegraphics[height=3cm]{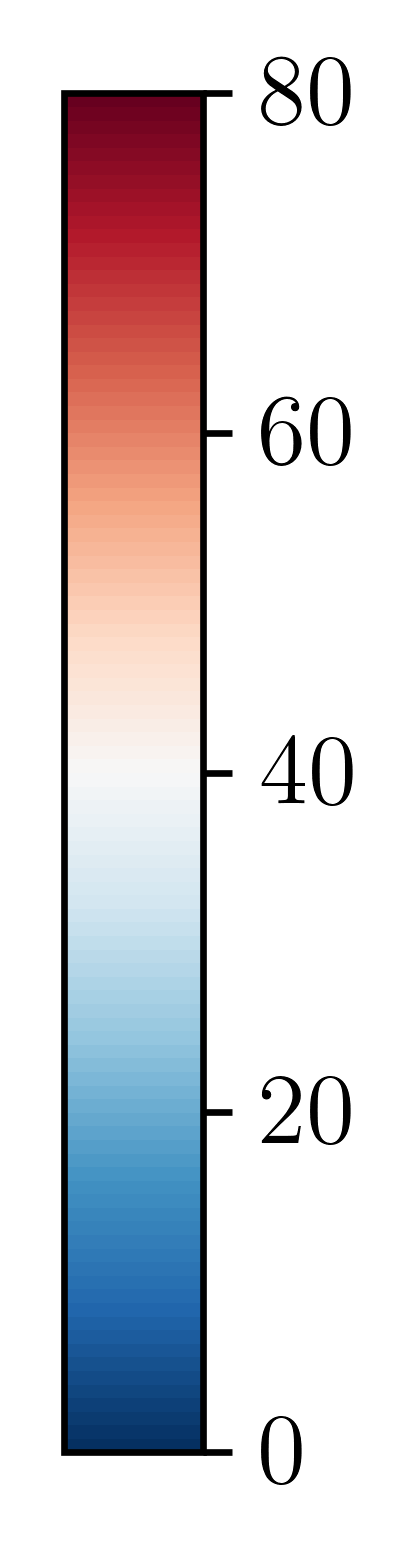}
        \vspace{0cm}
    \end{subfigure}
    ~
    \begin{subfigure}[t]{0.25\textwidth}
        \centering
        % \vspace{-3cm}
        \includegraphics[height=3cm]{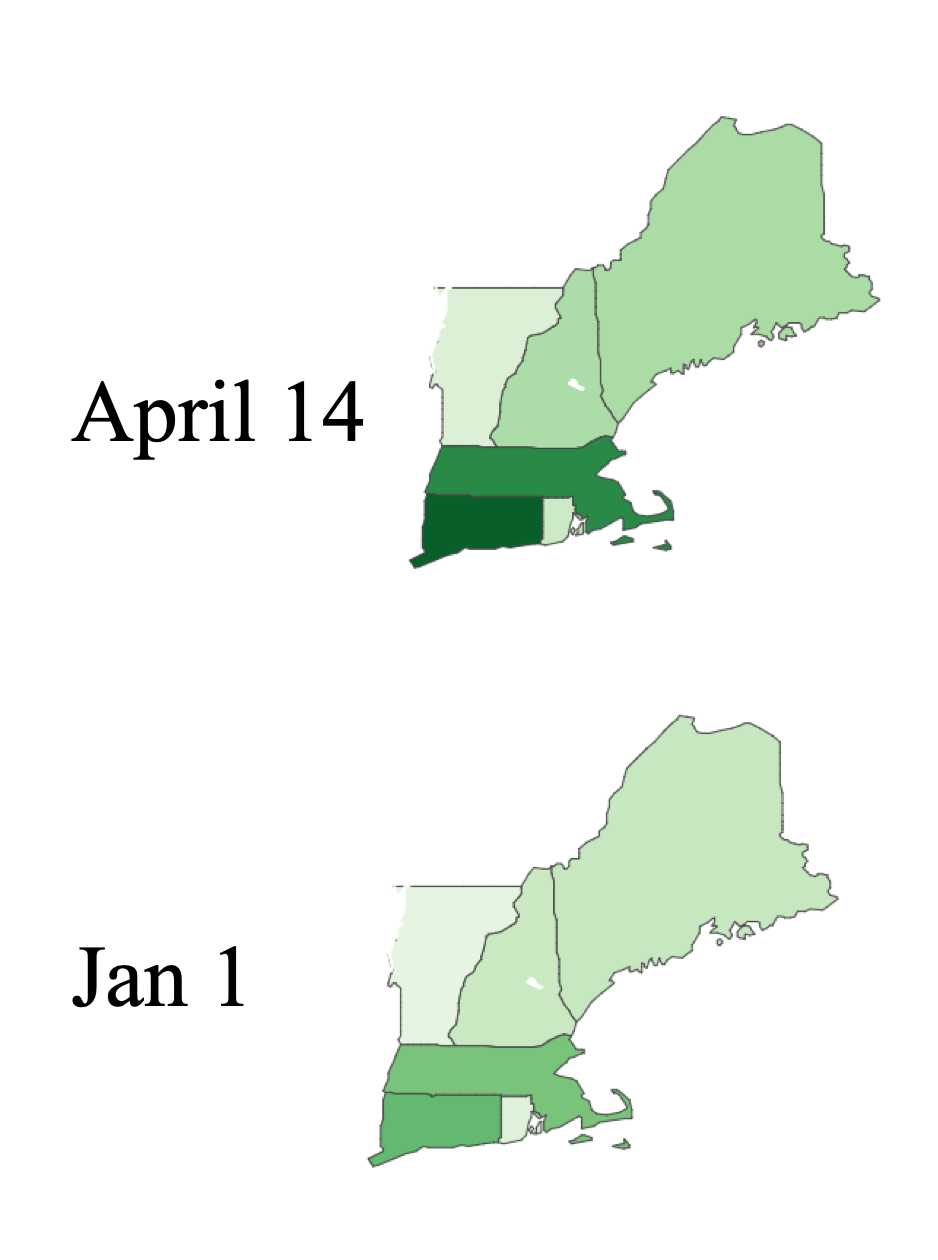}
        ~
        \includegraphics[height=3cm]{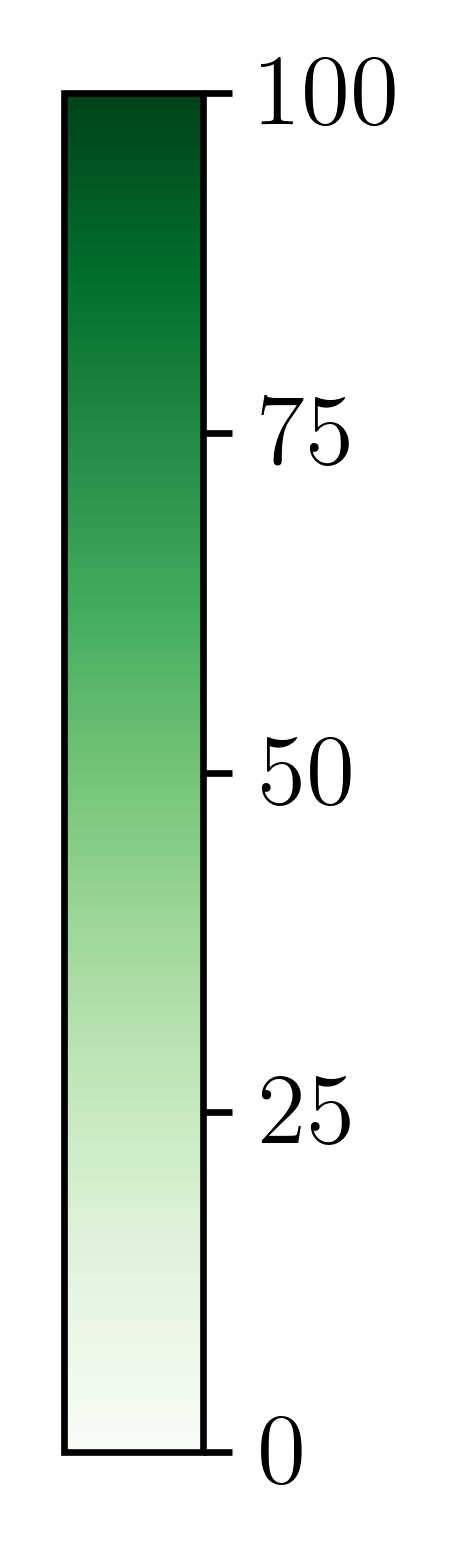}
        \caption{Energy consumption (GWh)}
    \end{subfigure}%
    % ~
    % \begin{subfigure}[t]{0.05\textwidth}
    %     \centering
    %     \includegraphics[height=3cm]{figs/states_colorbar.png}
    %     \vspace{0cm}
    % \end{subfigure}
    
    \begin{subfigure}[b]{0.95\textwidth}
        \centering
        \includegraphics[height=4cm]{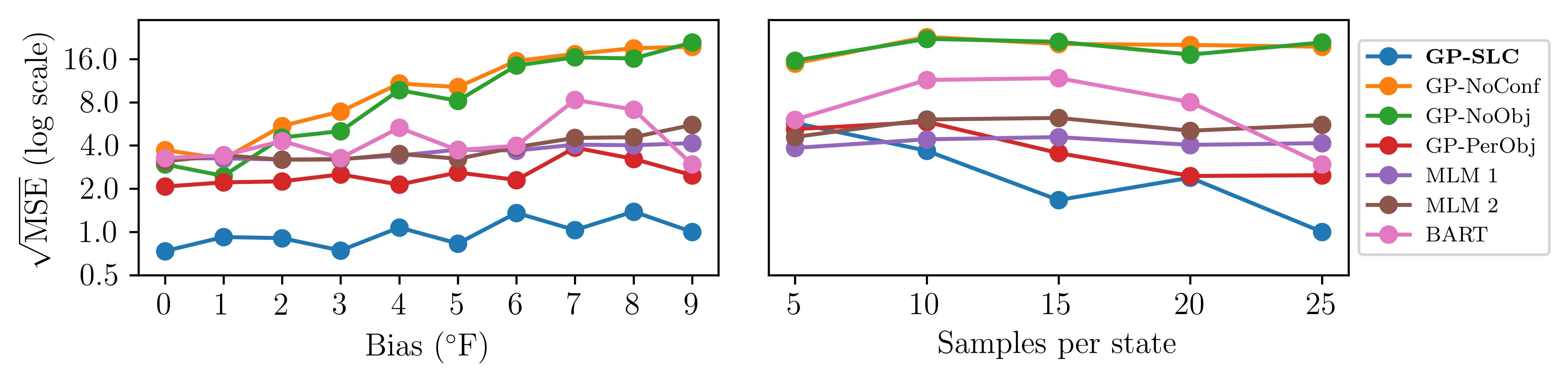}
    \caption{Mean squared error in estimated sample average treatment effect.}
    \label{fig:bias_results}
    \end{subfigure}
    
    \caption{\textbf{Process and results for New England energy consumption benchmark.} We sample hotter days with higher probability for states with higher daily energy consumption (a-d). Sampling in this way simulates confounding, creating an observational relationship (consumption is signicantly higher in hotter days) that differs from the causal relationship (low or high temperature causes a moderate increase in energy consumption). GP-SLC (this paper) produces accurate estimates of counterfactual outcomes, despite this confounding bias (e). For baselines that ignore hierarchical structure (GP-NoObj and GP-NoConf), accuracy decreases significantly with increasing confounding bias. Results are normalized by the $\sqrt{\textrm{MSE}}$ of the GP-SLC model with $\textrm{bias} = 9^{\circ}F$ and 25 samples per state.}
    \label{fig:bias}
\end{figure*}

\section{Experiments}
\label{sec:Experiments}

Unlike associational models, which can be evaluated using accuracy on held-out test data, causal models produce predictions about unobserved \textit{counterfactual} distributions. As a result, effective evaluation of causal models requires different methods~\cite{gentzel2019case}. We evaluate the GP-SLC model using three benchmarks with known counterfactual outcomes. In Section~\ref{subsec:synthetic}, we evaluate GP-SLC using a fully synthetic hierarchical data generating process. In Section~\ref{subsec:IHDP} we modify the Infant Health and Development Program (IHDP) benchmark~\cite{hill2011bayesian} to include hierarchical structure and latent confounders. In Section~\ref{subsec:ISO} we introduce and evaluate on a new benchmark task for observational causal inference with hierarchical data, predicting the effect of changes in temperature on state-wide electric energy consumption in New England (NEEC).

We implement the GP-SLC model using Gen~\cite{cusumano2019gen}, a probabilistic programming language with programmable inference. Except where otherwise specified we set $N_U=3$ and $\alpha_{\theta}=\beta_{\theta}=4$ for each inverse gamma prior over kernel hyperparameters and exogenous noise variance. We estimate individual treatment effects using Algorithm~\ref{alg:inference}, with $\textrm{N}_{\textrm{Outer}}=5000$, $\textrm{N}_\textrm{MH}=3$, $\textrm{N}_{\textrm{ES}}=5$, and $\textrm{drift}_{\theta} = 0.5, \forall \theta \in \Theta$.

We compare the GP-SLC model against six baselines: a GP regression model that ignores latent confounding variables (GP-NoConf), a GP-SLC model where each instance is incorrectly assigned a single object (GP-NoObj), a seperate GP regression model for each object (GP-PerObj), Bayesian additive regression trees (BART)~\cite{hill2011bayesian}, a random slope and intercepts linear model (MLM 1), and a random intercepts linear model (MLM 2)~\cite{gelman2006multilevel}. The Gaussian process baselines are ablations of the full GP-SLC model, and use the same kernels, priors over hyperparameters, and inference scheme. The BART baseline uses the object identifier, $o$, as an additional covariate. See the supplementary materials for additional details on baselines.

We use two evaluation metrics to evaluate GP-SLC and baselines, \textit{mean squared error of the sample average treatment effect}, $\text{MSE} = \mathbb{E}_{t_*}[(\SATE^*_{t_*} - \SATE_{t_*})^2]$, and \textit{precision in estimation of heterogenous effect}~\cite{hill2011bayesian}, $\text{PEHE} = \mathbb{E}_{t_*}[\sum_i^{N_i} (\ITE^*_{i, t_*} - \ITE_{i, t_*})^2 /N_i]$, where $\ITE^*_{i, t_*}$ and $\SATE^*_{T_*}$ are the actual effects and $\ITE_{i, t_*}$ and $\SATE_{t_*}$ are the predicted effects. For the synthetic benchmark, we average over 100 regular intervals between the 5th and 95th percentile of treatment assignment in the observational data. For the NEEC benchmask, we average over $\{30, 30.1, ..., 70^\circ{F}\}$.

\begin{figure*}[t!]
    \centering
    \includegraphics[width=\textwidth]{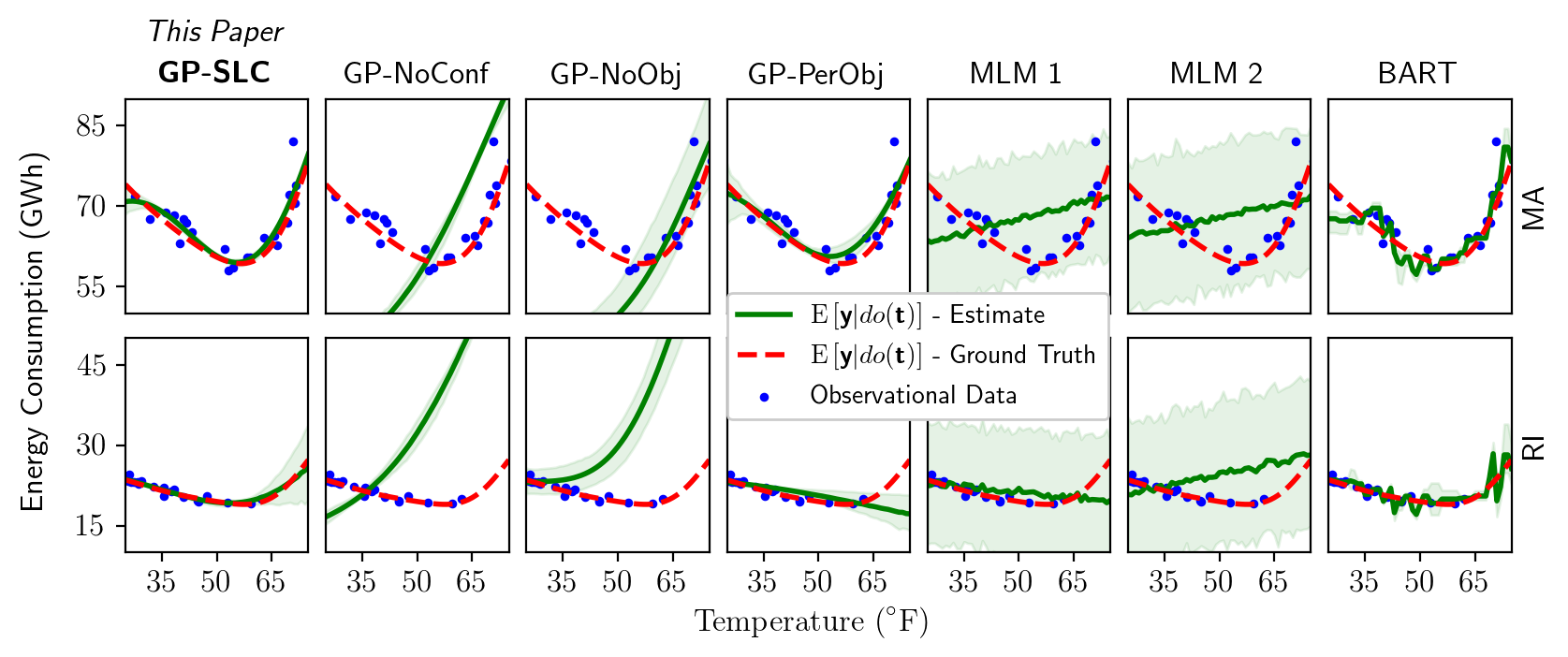}
    \caption{\textbf{Comparison among methods on the New England energy consumption benchmark.} Above are GP-SLC and all baselines' effect estimates on the NEEC benchmark with $\text{bias} = 9^{\circ}F$ and 25 samples per state. Green shaded regions indicate $90\%$ credible intervals. GP-SLC effectively recovers the effect of temperature on energy consumption, despite the latent confounding introduced by biased sampling. The best performing baseline, GP-PerObj, produces poor estimates of the effect of high temperatures in Rhode Island.}
    \label{fig:NEECresults}
\end{figure*}

\subsection{Synthetic Data}
\label{subsec:synthetic}

We evaluate GP-SLC and various baselines on two synthetic datasets with hierarchically structured latent confounders, one with additive and one with multiplicative treatment and outcome functions. Both synthetic datasets are generated using three dimensional object-level confounders for 20 objects, each of which contains 10 instances. Observed instance-level covariates are generated as a linear function of object-level Gaussian distributed latent confounders. Details for synthetic treatment and outcome functions are presented in the supplementary materials, and evaluation results are shown in Table~\ref{tab:synthetic}. GP-SLC consistently matches and exceeds the counterfactual prediction performance of the six baselines on synthetic data. Baselines that ignore object structure (GP-NoConf and GP-NoObj) produce the least accurate counterfactual predictions.

\begin{table}[b!]
  \begin{tabular}{r|cc|cc}
    \toprule
    \multirow{2}{*}{Model}  &
      \multicolumn{2}{c}{Additive} &
      \multicolumn{2}{c}{Multiplicative}\\
      & {$\sqrt{\text{PEHE}}$} & {$\sqrt{\text{MSE}}$} & {$\sqrt{\text{PEHE}}$} & {$\sqrt{\text{MSE}}$}\\
      \midrule
      \textbf{GP-SLC} & \textbf{1.0} & \textbf{1.0} & \textbf{1.0} & 1.0 \\
      GP-NoConf & 21.3 & 25.3 & 4.2 & 7.6 \\
      GP-NoObj & 22.2 & 27.0 & 4.5 & 8.1\\
      GP-PerObj & 3.7 & 3.4 & 1.1 & \textbf{0.9}\\
      MLM1 & 1.2 & 1.02 & 2.4 & 2.9 \\
      MLM2 & 1.3 & 1.6 & 4.4 & 9.3 \\
      BART & 8.5 & 10.7 & 2.6 & 4.3\\
    \bottomrule
  \end{tabular}
  \caption{Results on synthetic data with additive and multiplicative nonlinear treatment and outcome functions. Scores are normalized by the score of GP-SLC. Lower is better.} \label{tab:synthetic}
\end{table}
% \newpage

In addition to the synthetic experiments presented in Table~\ref{tab:synthetic}, we tested the behaviour of GP-SLC using two alternative synthetic data generating processes. On the first, a linear structural data generating process with shared confounding, GP-SLC produces comparable estimates to the multi-level model baselines. On the second, in which each object shares a common effect of treatment and outcome rather than a common cause, GP-SLC is not susceptible to collider bias~\cite{berkson1946limitations,elwert2014endogenous}. This empirical finding is consistent with recent theory on object conditioning~\cite{jensen2019object}.

\subsection{Infant Health and Development Program}
\label{subsec:IHDP}

The IHDP benchmark~\cite{hill2011bayesian} uses real data for treatments (whether a child recieves high-quality child care and home visits from a trained provider) and covariates (birth weight, head circumference, etc.) from the 1992 Infant Health and Development Program~\cite{ramey1992infant} with a synthetic nonlinear outcome function. We modify the IHDP benchmark to simulate hierarchically structured data by randomly duplicating 30\% of the data instances and reassigning the duplicate's treatment assignment to be the opposite of the original instance. In order to introduce variation between duplicated instances, we add noise to each individuals' continuous covariates from a $\mathcal{N}(0, \sigma^2_j)$, where $\sigma^2_j$ is 5\% of the $j$th covariate's marginal variance. We obscure the remaining 15 categorical covariates, representing object-level latent confounding. Even though the 15 categorical covariates are obscured from the GP-SLC model, they are identical across duplicates, unlike the observed covariates. We then generate observed and counterfactual outcomes using the benchmark synthetic outcome function, applied to treatment, modified covariates, and latent confounders. In this setting, $Pa(i)=Pa(i')$ if instance $i$ is a duplicate of instance $i'$ or vice versa. Although each duplicate's treatment assignment is deterministic, the overall relationship between treatment and outcome is still confounded, as we only duplicate a subset of the original instances. 

For the IHDP benchmark, which has binary treatment variables, we modify the GP-SLC model by replacing the expression $\T_{i} = f_t(\U_{o=Pa(i)}, \X_i, \eps_{t_i})$ with the expressions $\hat{\T}_{i} = f_{\hat{\T}}(\U_{o=Pa(i}), X_i, \eps_{\hat{t}_i})$ and $\T_{i} \sim \text{Bernoulli}(\textit{expit}(\hat{\T}_i))$. In this setting, we use elliptical slice sampling to approximate the latent logit probability of treatment, $\hat{\T}$.

Given the small size of each object, we omit the GP-PerObj baseline model from this evaluation. As the IHDP benchmark includes binary treatment variables we compared against four additional baselines: balanced linear regression (BalReg) and balanced neural nets (BALNN)~\cite{Johansson2016LearningRF}, targeted maximum likelihood estimation with the superlearner (TMLE)~\cite{van2007super}, and inverse probability of treatment weighting with logistic regression (IPTW)~\cite{imbens2015causal}. 

Results of the IHDP evaluation are presented in Table~\ref{tab:IHDP}. GP-SLC matches and exceeds the performance of other baselines when predicting the effect of assigning treatment to individuals who were previously untreated. In this setting, the linear models (MLM 1 and MLM 2) produce the least accurate counterfactual predictions.

\begin{table}[t!]
  \begin{tabular}{r|cc|cc}
    \toprule
    \multirow{2}{*}{Model}  &
      \multicolumn{2}{c}{Control} &
      \multicolumn{2}{c}{Treated}\\
      & {$\sqrt{\text{PEHE}}$} & {$\sqrt{\text{MSE}}$} & {$\sqrt{\text{PEHE}}$} & {$\sqrt{\text{MSE}}$}\\
      \midrule
      \textbf{GP-SLC} & \textbf{1.0} & \textbf{1.0} & 1.0 & 1.0 \\
      GP-NoConf & 1.03 & 1.07 & 1.04 & 0.94 \\
      GP-NoObj & 1.11 & 1.02 & \textbf{0.82} & 1.08 \\
      MLM1 & 68.3 & 33.2 & 106.7 & 1028.4 \\
      MLM2 & 73.3 & 389.1 & 45.8 & 63.2 \\
      BART & 3.7 & 1.1 & 2.4 & \textbf{0.33}\\
      BALReg & 5.1 & 82.7 & 1.9 & 0.5 \\
      BALNN & 2.1 & 7.0 & 1.7 & 4.5 \\
      TMLE & n/a & 209.8 & n/a & 12.2 \\
      IPTW & n/a & 50.6 & n/a & 90.5 \\
    \bottomrule
  \end{tabular}
  \caption{Results on the modified infant health and development program benchmark, shown seperately for treated and untreated individuals. Scores are normalized by the score of GP-SLC. TMLE and IPTW do not estimate individual treatment effects. Lower is better.} \label{tab:IHDP}
\end{table}
% \newpage

\subsection{New England Energy Consumption}
\label{subsec:ISO}

We introduce a new benchmark for estimating heterogenous effects in hierarchically structured settings, predicting the effect of changing temperature on state-wide electric energy consumption in New England. Unlike the evaluation in Section~\ref{subsec:IHDP}, which includes real treatments, covariates, and confounders and a synthetic outcome function, the New England energy consumption (NEEC) benchmark preserves outcome functions from real quasi-experimental data, and uses biased sampling to induce confounding. Specifically, we generate data for the NEEC benchmark task using the New England Independent Service Operator's public records on hourly dry-bulb temperature and state-wide energy consumption for the 2018 calendar year~\cite{iso2018data}, which we then aggregate into daily averages. 

While the marginal distribution over daily average temperature is nearly identical across states in the original dataset, the causal relationship between temperature and energy consumption differs across states, likely due to differences in population density,  and commercial/industrial activity. To introduce confounding, we systematically sample days (instances) from states (objects) based on the state's typical energy consumption, including hotter days with higher probability for high consuming states. Specifically, we use importance resampling with a target distribution over Farenheit temperatures $T\sim \mathcal{N}(45 + bias \cdot s_{o}, 15)$, where $s_{CT}=3, s_{MA}=2, s_{ME}=1, s_{NH}=-1, s_{RI}=-2, s_{VT}=-3$. An example of this sampling with $\text{bias}=9$ is shown in Figure~\ref{fig:bias} (a-c). Biased sampling in this way introduces a statistical dependency across the dataset (consumption is significantly higher in hotter days), that differs from the causal relationship (low or high temperature causes a moderate increase in energy consumption). This approach of sampling quasi-experimental data to simulate confounding is an emerging standard in causal inference evaluation~\cite{gentzel2019case} although existing benchmarks are not hierarchically structured. Figure~\ref{fig:bias} (a-d) shows an example of this sampling process for the NEEC benchmark.

\begin{table}[b!]
    \centering
  \begin{tabular}{r|c|c|c|c|c|c}
    \toprule
    Model  & CT & MA & ME & NH & RI & VT \\
    \midrule
      \textbf{GP-SLC} & \textbf{1.0} & \textbf{1.0} & \textbf{1.0} & \textbf{1.0} & \textbf{1.0} & 1.0 \\
      GP-NoConf & 13.2 & 13 & 31.5 & 41.6 & 47.4 & 14.9 \\
      GP-NoObj & 19.1 & 14 & 26.8 & 36.2 & 48 & 16.5 \\
      GP-PerObj & 1.6 & 1.3 & 5.2 & 9.7 & 6.5 & \textbf{0.7} \\
      MLM1 & 6.9 & 5 & 25.0 & 5 & 5.1 & \textbf{0.7} \\
      MLM2 & 6.4 & 4.9 & 39.4 & 6.3 & 9.9 & 3.8 \\
      BART & 4.1 & 2.1 & 13.3 & 3.6 & 3.3 & 2.4 \\
    \bottomrule
  \end{tabular}
  
  \caption{$\sqrt{\textrm{MSE}}$ for the New England energy consumption benchmark, with $\text{bias} = 9^{\circ}F$ and 25 samples per state. Lower is better. Scores are normalized by GP-SLC's score for the same state.} \label{tab:NEEC}
\end{table}
\raggedbottom

Sampling in this way does not provide instance-level counterfactual outcomes. Instead, we estimate the sample-average ground truth counterfactual outcome by fitting a Gaussian process regression model for each state, using treatments and outcomes from the entire calendar year.

Figure~\ref{fig:bias_results} shows the models' performances with varying degree of confounding and sample sizes, and Figure~\ref{fig:NEECresults} shows the estimated and actual effect of temperate on electric energy consumption for two of the six states. Despite the induced confounding, GP-SLC consistently produces accurate estimates of causal effect. The baselines that ignore confounding (GP-NoConf and GP-NoObj) perform poorly as the degree of confounding increases, incorrectly attributing sample-wide association as indicative of causal effect. The linear multi-level models (MLM 1 and MLM 2) are not biased by confounding, but produce poor estimates due to their restrictive parametric assumptions. The remaining two baselines (GP-PerObj and BART) produce more accurate estimates than the other four baselines, but still overfit.

\subsection{Limitations}

Despite the fact that GP-SLC produces state-of-the-art counterfactual predictions on most of our synthetic and semisynthetic benchmarks, it tends to underestimate the uncertainty in these estimates. In other words, the posterior density on the ground-truth counterfactual is sometimes low, despite the fact that the mean estimate is close to the ground-truth relative to the baselines. We suspect that this is partially attributable to inaccuracies resulting from our approximate inference procedure (Algorithms \ref{alg:inference_hyp} and \ref{alg:inference_conf}). Alternative approximate inference schemes, such as using our current approach as a rejuvenation move in a sequential Monte Carlo (SMC) algorithm~\cite{doucet2001introduction}, may resolve these inaccuracies. This kind of SMC-based inference procedure may also help GP-SLC scale to problems with more covariates and objects than we explore in this paper.

Our empirical study focusses on data generating processes that satisfy GP-SLC's implicit semiparametric assumptions; (i) covariates for individuals belonging to the same object are marginally Gaussian distributed, and (ii) exogenous noise is additive and Gaussian. The effect of these modeling assumptions on counterfactual prediction and estimates of effect strength needs additional empirical characterization, ideally via large-scale synthetic experiments (where ground truth is known and robustness to modeling bias can be qualitatively studied).

\section{Related Work}

Leveraging hierarchical structure is well-established as a technique for adjusting for latent confounding~\cite{gelman2006multilevel, gelman2006data, hong2006evaluating}. Using Gaussian processes for causal inference is also well-established~\cite{alaa2017bayesian, alaa2018bayesian, silva2010gaussian, schulam2017reliable, Zhang2010InvariantGP}, as is the use of generative model approaches to adjust for latent confounders given restrictions on structure~\cite{miao2018proxy, louizos2017causal, tran2018implicit, wang2019blessings}. To the best of our knowledge, GP-SLC is the first semiparametric generative modeling approach that leverages hierarchical structure to adjust for latent confounders. 

GP-SLC is one of many recent techniques~\cite{Shalit2017GeneralizationandBounds, Johansson2016LearningRF} for estimating individual-level treatment effects. Prior work focusses on the propositional setting under \textit{strong ignorability}, i.e. with no latent confounders. We focus on the hierarchical setting in which latent confounders are shared across multiple instances.

Recent work~\cite{schulam2017reliable} has used Gaussian process models for causal inference in temporal settings, which assumes unconfoundedness and that the outcome is smooth with respect to time and covariates. GP-SLC allows for the existence of object-level latent confounders, and instead assumes that the outcome is smooth with respect to treatment assignment, covariates, and latent confounders. Longitudinal data analysis is closely related to the hierarchical settings we consider in this work: measurements (instances) of individuals (objects) are repeated over a period of time. Extending GP-SLC to the setting where latent confounders are not shared across instances, but instead change over time, is an exciting area of future work.

GP-SLC is most similar to~\cite{alaa2017bayesian}, in that their approach also uses GP models to estimate individual treatment effects. However, GP-SLC: (i) handles hierarchical latent confounders by first performing inference over object-level latent variables; (ii) accounts for the covariance between noise-free factual and counterfactual outcomes (see $\Sigma_{12}$ and $\Sigma_{21}$ in the Supplementary materials); and (iii) uses a Monte Carlo algorithm for inference that yields quantified uncertainty estimates. Their approach could be applied in hierarchical settings by treating the object identifier $o$ as a categorical covariate and using a delta kernel to construct the outcome kernel covariance matrix. This is identical to the GP-PerObj baseline, except that GP-PerObj does not share inferred kernel hyperparameters across objects.

\section{Conclusions}

This paper presents GP-SLC, a Gaussian process model for causal inference with hierarchically structured latent confounders. In Section~\ref{sec:Experiments}, we show that, compared to widely used alternatives, GP-SLC produces more accurate estimates of causal effect in realistic sparse observational settings where strong prior knowledge about structure can inform causal estimates. The hierarchical structure we exploit in this paper is one of many kinds of structural background knowledge that could improve causal estimates, and developing techniques to exploit such knowledge is an important area of future work. Extending GP-SLC to handle large observational datasets~\cite{cao2018scaling, quinonero2005unifying} or to leverage experimental evidence~\cite{witty2019bayesian} are also exciting areas of future work.

\newpage
\section*{Acknowledgments}
Thanks to Marco Cusumano-Towner, Feras Saad, Alex Lew, Cameron Freer, Rachel Paiste, Amanda Gentzel, Andy Zane, Jameson Quinn, and the anonymous reviewers for their helpful feedback and suggestions. Sam Witty, Kenta Takatsu, and David Jensen were supported by DARPA and the United States Air Force under the XAI (Contract No. HR001120C0031) and CAML (Contract No. FA8750-17-C-0120) programs, respectively. Vikash Mansinghka was supported by DARPA under the SD2 program (Contract No. FA8750-17-C-0239) and a philanthropic gift from the Aphorism Foundation. Any opinions, findings and conclusions or recommendations expressed in  this material are those of the authors and do not necessarily reflect the views of DARPA or the United States Air Force. 

% \newpage

\bibliography{ref.bib}
\bibliographystyle{icml2020}

\onecolumn
\section{Supplementary Materials}
\subsection{Kernel Functions}

In this section, we present a detailed definition of each of the kernel functions used in GP-SLC: 

\begin{align*}
    k'_{x_k}([\U_{o=Pa(i)}], [\U_{o'=Pa(i')}]) &= \sigma^2_{x} \exp{\left[-\sum_j^{N_U}\frac{(\U_{o, j} - \U_{o', j})^2}{\lambda_{ux_{j,k}}}\right]}\\
    k'_{t}([\U_{o=Pa(i)}, \X_i], [\U_{o'=Pa(i')}, \X_{i'}]) &= \sigma^2_{t} \exp{\left[-\sum_j^{N_U} \frac{(\U_{o, j} - \U_{o', j})^2}{\lambda_{ut_j}} - \sum_k^{N_X} \frac{(\X_{i, k} - \X_{i', k})^2}{\lambda_{xt_{k}}}\right]}\\
    k'_{y}([\U_{o=Pa(i)}, \X_i, \T_i], [\U_{o'=Pa(i')}, \X_{i'}, \T_{i'}]) &= \sigma^2_{y} \exp{\left[-\sum_j^{N_U} \frac{(\U_{o, j} - \U_{o',j})^2}{\lambda_{uy_j}} - \sum_k^{N_X} \frac{(\X_{i, k} - \X_{i', k})^2}{\lambda_{xy_{k}}} - \frac{(\T_i - \T_{i'})^2}{\lambda_{ty}}\right]}.
\end{align*}
where $\lambda_{*}$ is a lengthscale hyperparameter and defined for each dimension of corresponding variables. Here, each dimension of $\X$ is generated independently given $\U$, and $k'_{x_k}$ refers to the kernel function for the $k$th dimension of $x$. Intuitively, each kernel lengthscale determines the relative strength of influence of each variable's parents in Equation 1. For example, if $\lambda_{ty} >> \lambda_{xy_{i=1...N_X}}$, the covariance between instances (or counterfactuals) with similar treatments will be greater than the covariance between instances with similar covariates. 

\subsection{Exact Inference: $\boldsymbol{Y'_*} -\boldsymbol{Y'}$ Details}

Here we provide additional details on how to compute GP-SLC's conditional distribution over individual treatment effects. Given the expression for $\Bigg(\begin{bmatrix*}[l]Y \\Y' \\ Y'_*\end{bmatrix*}\Bigg|\;T_*, T, X, U, \Theta \Bigg)$ in Section 4.2, conditioning on $Y$ yields the following:

\begin{equation*}
    \Big(\begin{bmatrix*}[l]Y' \\ Y'_*\end{bmatrix*}\Big|T_*, Y, T, X, U, \Theta \Big) \sim \mathcal{N}\Big(\begin{bmatrix} \mu_1 \\ \mu_2 \end{bmatrix}, \begin{bmatrix} \Sigma_{1,1} & \Sigma_{1,2} \\ \Sigma_{2,1} & \Sigma_{2,2} \end{bmatrix}\Big)
\end{equation*}
where, 
\begin{align*}
    \mu_1 &= \K'(\W, \W) \inv{\K(\W, \W)}Y &
    \mu_2 &= \K'(\W, \W_*) \inv{\K(\W, \W)}Y\\
    \Sigma_{1,1} &=  \K'(\W, \W) - \K'(\W, \W)\inv{\K(\W, \W)}\K'(\W, \W) &
    \Sigma_{1,2} &= \K'(W, W_*) - \K'(W, W)\inv{\K(W, W)}\K'(W, W_*)\\
    \Sigma_{2,1} &= \K'(W_*, W) - \K'(W_*, W)\inv{\K(W, W)}\K'(W, W) &
    \Sigma_{2,2} &= \K'(W_*, W_*) - \K'(W_*, W)\inv{\K(W, W)}\K(W, W_*)
\end{align*}

As the difference of variables that are jointly Gaussian is Gaussian, we have that $(Y'_*-Y'|T_*, X, T, Y, U, \Theta) \sim \mathcal{N}(\mu_{\ITE}, \Sigma_{\ITE})$, where $\mu_{\ITE} = \mu_2 - \mu_1$ and $\Sigma_{\ITE} = \Sigma_{1,1} - \Sigma_{1,2} - \Sigma_{2,1} + \Sigma_{2,2}$.

\subsection{Asymptotic Posterior Consistency}

Here we provide proofs for  Proposition 5.1 and Theorems 5.2 and 5.3. The analysis in this section follows the setup presented in \cite{d2019multi}, with the inclusion of shared latent confounding amongst individual instances. We omit covariates $X$ from this analysis and assume that $N_U=1$ for brevity without loss of generality. Note that these theoretical results also hold for the random intercepts multilevel model~\cite{gelman2006multilevel}.

% \subsubsection{Setup}

Assuming linear kernels and additive Gaussian exogenous noise, we can equivalently rewrite the GP-SLC model as follows. This equivalent structural causal model is parameterized by latent variables $\alpha, \beta, \tau \in \mathbb{R}$ and $\sigma^2_U, \sigma^2_T, \sigma^2_Y \in \mathbb{R^+}$. For all $o \in 1,...,N_O$ and $i \in 1,...,N_I$, we have that:

\begin{equation*}
    \begin{aligned}
        \eps_{u_{o}} &\sim \mathcal{N}(0, \sigma^2_U)\\
        \eps_{t_{i}} &\sim \mathcal{N}(0, \sigma^2_T) \\
        \eps_{y_{i}} &\sim \mathcal{N}(0, \sigma^2_Y)
    \end{aligned}
    \quad 
    \begin{aligned}
        \U_{o} &= \eps_{u_o}\\
        \T_{i} &= \alpha \U_{o=Pa(i)} + \eps_{t_i}\\
        \Y_{i} &= \beta \T_i + \tau \U_{o=Pa(i)} + \eps_{y_i}.
    \end{aligned}
    \label{eq:GP-SLC-lin}
\end{equation*}

In this setting, estimating individual treatment effect reduces to estimating $\beta$, as $\Y_{i, \T_*} - \Y_i = \beta (\T_* - \T_i)$. We make the following observations.

% \begin{prop}
\textbf{Proposition 5.1} \textit{When $N_O = N_I$, $\ITE_{t_*}$ is not asymptotically consistent $\forall t_* \in \mathbb{R}$.} 
% \label{prop:diag}
% \end{prop}

For a detailed proof of Proposition 5.1, see Proposition 1 in \cite{d2019multi}. In summary, they show that given any set of latent parameters $\Theta = (\alpha, \beta, \tau, \sigma^2_U, \sigma^2_T, \sigma^2_Y)$, there exists an alternative set of parameters $\Theta'$ such that $P(T, Y|\Theta) = P(T, Y|\Theta')$ and $\beta \neq \beta'$. In other words, the structural causal model forms a linear system of equations that is rank-deficient. The set of parameters that satisfy this condition construct an \textit{ignorance region}.

Extending their results to the Bayesian setting, we have that for any two sets of parameters $\Theta$ and $\Theta'$ on the same ignorance region, the posterior odds ratio reduces to the prior odds ratio, $\frac{P(\Theta|T,Y)}{P(\Theta'|T,Y)} =\frac{P(\Theta)P(T, Y|\Theta)}{P(\Theta')P(T, Y|\Theta')} = \frac{P(\Theta)}{P(\Theta')}$. By definition, $\Theta$ is not asymptotically consistent, as the posterior $P(\Theta|T, Y)$ depends on the prior $P(\Theta)$. The problem of \textit{asymptotic consistency} can be mitigated when $N_O < N_I$.

\textbf{Theorem 5.2} \textit{Assume there exists an object $o$ that is the parent of $n$ instances, $I' = \{i'_1, ..., i'_n\}$. Then $\ITE_{t_*}$ is asymptotically consistent as $n$ approaches $\infty, \forall t_* \in \mathbb{R}$.}

\begin{proof} 
For all $i' \in I'$, we have that $\Y_{i'} = \beta \T_{i'} + C + \eps_{y_{i'}}$ for some constant $C \in \mathbb{R}$. Therefore, the covariance between $T$ and $Y$ in $I'$ is uniquely given by $\beta$, i.e. $cov(\T_{i'\in I'},\Y_{i'\in I'}) = \beta$. Estimating the covariance of a bivariate normal has a unique maximum likelihood solution. Therefore, by the Bernstein-von Mises Theorem~\cite{doob1949application} we have that the posterior over $\beta$, and thus $\ITE_{t_*}$, is asymptotically consistent as $n$ approach $\infty$.
\end{proof}

% \begin{theorem}
% \label{thm:manyblocks}
\textbf{Theorem 5.3}\textit{
Assume there exists $n$ objects $\mathbb{O} = \{o_1, ..., o_n\}$, each of which are the unique parents of $k \geq 2$ instances $I'_o = \{i'_{o,1}, ..., i'_{o, k_o}\}$. Then $\ITE_{t_*}$ is asymptotically consistent as $n$ approaches $\infty$.}
% \end{theorem}

\begin{proof} For all $o\in \mathbb{O}$, $j \in \{1,...,k_o\}$ let  $\T'_{i'_{o,j}} = \T_{i'_{o,j}} - \bar{\T}_o$ and $\Y'_{i'_{o,j}} = \Y_{i'_{o,j}} - \bar{\Y}_o$, where $\bar{\T}_o = \sum_j {\T_{i'_{o,j}}}/k_o$ and $\bar{\Y}_o = \sum_j{\Y_{i'_{o,j}}}/k_o$, i.e., the sample average over all instances that share a parent object. Therefore, $\T'_{i'_{o,j}} = \alpha \U_o + \eps_{t_{i'_{o,j}}} - \sum_j (\alpha \U_o + \eps_{t_{i'_{o,j}}} )/k_o = \eps_{t_{i'_{o,j}}} - \sum_j {\eps_{t_{i'_{o,j}}}}/k_o$ and $\Y'_{i'_{o,j}} = \beta (\alpha \U_o + \eps_{t_{i'_{o,j}}}) + \tau \U_o + \eps_{y_{i'_{o,j}}} - \sum_j(\beta (\alpha \U_o + \eps_{t_{i'_{o,j}}}) + \tau \U_o + \eps_{y_{i'_{o,j}}})/k_o = \beta \T'_{i'_{o,j}} + \eps_{y_{i'_{o,j}}} - \sum_i {\eps_{y_{i'_{o,j}}}}/k_o$. As $\eps_{y_{i'_{o,j}}}$ is independent of $\T'_{i'_{o,j}}$, we have that the covariance between $\T'_{i'_{o,j}}$ and $\Y'_{i'_{o,j}}$ is equal to $\beta$. Therefore, the problem of estimating $\beta$ reduces to estimating the covariance of a bivariate normal distribution, $P(T', Y')$, which has a unique maximum likelihood solution. As in the proof of Theorem 5.2, by the Bernstein-von Mises Theorem~\cite{doob1949application} we have that the estimate of $\beta$, and thus $\ITE_{t_*}$, is asymptotically consistent as $n$ approach $\infty$.
\end{proof}

\subsection{Bayesian Linear Multilevel Model Baseline}

One of the baselines we use in the experiments is Bayesian linear multilevel models~\cite{gelman2006multilevel}. We implement two multilevel models, which introduce varying degrees of shared parameters across objects. The first multilevel model, also known as a random slope and intercepts model, (MLM 1) fits the observations using the following structural equations. 

\begin{gather*}
    \sigma_y^2 \sim \inv{\gamma}(\alpha_{\sigma_y}, \beta_{\sigma_y})\\
    \alpha \sim \mathcal{N}(\mu_\alpha, \Sigma_\alpha)\\
    \beta_o \sim \mathcal{N}(\mu_{\beta}, \sigma^2_{\beta}) \text{ for $o = 1 \dots N_O$}\\
    \eta_o \sim \mathcal{N}(\mu_{\eta}, \sigma^2_{\eta}) \text{ for $o = 1 \dots N_O$}\\
    \Y_i \sim \mathcal{N}(\beta_{o=Pa(i)} \T_i + \alpha^T \X_i + \eta_{o=Pa(i)}, \sigma_y^2)
\end{gather*}
This model allows varying intercepts $\eta$ and treatment effect $\beta$ across objects while assuming $\alpha$ is held constant across objects. 

The second multilevel model, also known as the random intercepts model, (MLM 2) fits the observations using the following structural equations. 

\begin{gather*}
    \sigma_y^2 \sim \inv{\gamma}(\alpha_{\sigma_y}, \beta_{\sigma_y})\\
    \alpha \sim \mathcal{N}(\mu_\alpha, \Sigma_\alpha)\\
    \beta \sim \mathcal{N}(\mu_\beta, \sigma^2_\beta)\\
    \eta_{o} \sim \mathcal{N}(\mu_{\eta}, \sigma^2_{\eta}) \text{ for $o = 1 \dots N_O$}\\
    \Y_i \sim \mathcal{N}(\beta \T_i + \alpha^T \X_i + \eta_{o=Pa(i)}, \sigma_y^2)
\end{gather*}

This model allows varying intercepts $\eta$ across objects while assuming $\alpha$ and $\beta$ are held constant across objects. 
We implement both models in Gen~\cite{cusumano2019gen}. For both models, we use $\alpha_{\sigma_y}=4.0, \beta_{\sigma_y}=4.0, \mu_{(\cdot)}=0,  \sigma^2_{\alpha}=3.0, \sigma^2_{\beta}=1.0,$ and $\sigma^2_{\eta}=10.0$ as priors.

\subsection{Synthetic Experiments}

We examine the finite-sample behavior of the GP-SLC model using two synthetic datasets that match GP-SLC's assumptions about the existence of \textit{object-level} latent confounders ($U$) that simultaneously influence \textit{instance-level} observed treatments (T), covariates ($X$), and outcomes ($Y$). The following structural equations summarize the data generating process: 
\begin{equation*}
\label{eq:synthetic_structural}
    \begin{aligned}
        W_{j} &\sim \mathcal{N}(0,1 I_3)\; \text{ for } j =  1, 2, 3 \\
        \U_{o} &\sim \mathcal{N}(0, 0.5I_3) \text{ for $o = 1 \dots N_O$}\\
        \X_i &= W \cdot \U_{o=pa(i)} + \boldsymbol{\epsilon}_{x_i} \text{ where } \boldsymbol{\epsilon}_{x_i} \sim \mathcal{N}(0, 0.5I_3) \text{ for } i = 1 ... N_I\\
        \T_i &= g_t(\X_i, \U_{o=pa(i)}) + \boldsymbol{\epsilon}_{t_i} \text{ where } \boldsymbol{\epsilon}_{t_i} \sim \mathcal{N}(0, 0.5)  \text{ for } i = 1 ... N_I\\
        \Y_i &= g_y(\T_i, \X_i, \U_{o=pa(i)}) + \boldsymbol{\epsilon}_{y_i} \text{ where } \boldsymbol{\epsilon}_{y_i} \sim \mathcal{N}(0,0.5)  \text{ for } i = 1 ... N_I
    \end{aligned}
\end{equation*}
First,  we draw $\U$ from a multivariate Gaussian distribution. Then, we generate covariates $\X$ as linear combinations of $\U$ with additive exogenous noise. We generate treatments $\T$ as a function ($g_t$) of $\X$ and $\U$  with additive noise. Finally, we generate outcome $\Y$ as a function ($g_y$) of $\X$, $\T$, and $\U$ with additive noise. For multi-dimensional variables, $\X$ and $\U$, we first apply the nonlinear function to each dimension of $\X$ and $\U$, then we aggregate them by summing across dimensions. 

The nonlinear treatment and outcome functions are shown in Table~\ref{tab:synth_funcs}. 

\begin{table*}[h!]
    \centering
    \begin{tabular}{c|c|c}
    \toprule
    Dataset   & $g_t(\X, \U)$ & $g_y(\T, \X, \U)$\\
    \midrule
    Additive    &  $\sum_{j}\X_{*,j} \; \text{sin}(\X_{*,j}) - \sum_{j}\U_{*,j}\; \text{sin}(\U_{*,j})$ & $\T\text{sin}(2\T) + \sum_{j}\X_{*,j}\;\text{sin}(\X_{*,j}) + 3\sum_{j}\U_{*,j}\;\text{sin}(\U_{*,j})$\\  
    Multiplicative  &  $\frac{1}{10}(\sum_{j}\X_{*,j}\;\text{sin}(\X_{*,j}))(\sum_{j}\U_{*,j}\;\text{sin}(\U_{*,j}))$ & $\frac{1}{10}(\T\text{sin}(2\T))(\sum_j\X_{*,j}\;\text{sin}(\X_{*,j}))(\sum_{j}\U_{*,j}\;\text{sin}(\U_{*,j}))$\\  
    \bottomrule
    \end{tabular}
    \caption{The functional form of $T$ and $Y$ for 2 synthetic datasets with continuous treatments and nonlinear outcome functions.} \label{tab:synth_funcs}
\end{table*}{}

\end{document}